\documentclass{article}

\usepackage{arxiv}

\usepackage[utf8]{inputenc} 
\usepackage[T1]{fontenc}    
\usepackage{hyperref}       
\usepackage{url}            
\usepackage{booktabs}       
\usepackage{amsfonts}       
\usepackage{nicefrac}       
\usepackage{microtype}      
\usepackage{lipsum}		
\usepackage{graphicx}
\usepackage{doi}
\usepackage{amsmath}
\usepackage{algorithm}
\usepackage{algpseudocode} 
\usepackage{amsmath}
\usepackage{tabularx}
\usepackage[T1]{fontenc}    
\usepackage{hyperref}       
\usepackage{url}            
\usepackage{booktabs}       
\usepackage{amsfonts}       
\usepackage{nicefrac}       
\usepackage{microtype}      
\usepackage{lipsum}
\usepackage{fancyhdr}       
\usepackage{graphicx}       
\graphicspath{{media/}}     

\usepackage{multirow}
\usepackage{makecell}

\title{All-Optical Doubly Resonant Cavities for ReLU Function in Nanophotonic Deep Learning}

\author{ \href{https://orcid.org/0000-0003-3372-6193}{\includegraphics[scale=0.06]{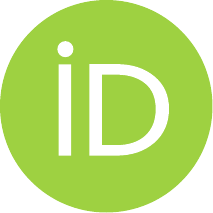}\hspace{1mm}Amirreza Ahmadnejad}\\
Department of Electrical Engineering\\
	Sharif University of Technology\\
	Tehran, Iran, 11155-4365 \\
\texttt{amirreza.ahmadnejad@sharif.edu} \\
	\And
	\href{https://orcid.org/0009-0005-1326-8377}{\includegraphics[scale=0.06]{orcid.pdf}\hspace{1mm}Mohmmad Mehrdad Asadi} \\
  Department of Electrical Engineering \\
	Sharif University of Technology\\
	Tehran, Iran, 11155-4365 \\
  \texttt{mehrdad.asadi79@sharif.edu} \\
	\And
	\href{https://orcid.org/0000-0002-3105-2511}{\includegraphics[scale=0.06]{orcid.pdf}\hspace{1mm}Somayyeh Koohi} \\
  Department of Computer Engineering \\
	Sharif University of Technology\\
	Tehran, Iran, 11155-4365 \\
  \texttt{koohi@sharif.edu } \\
}

\renewcommand{\headeright}{}
\renewcommand{\undertitle}{}
\renewcommand{\shorttitle}{All-Optical Doubly Resonant Cavities for Energy-Efficient ReLU Function in Nanophotonic Deep Learning}

\hypersetup{
pdftitle={A template for the arxiv style},
pdfsubject={q-bio.NC, q-bio.QM},
pdfauthor={David S.~Hippocampus, Elias D.~Striatum},
pdfkeywords={First keyword, Second keyword, More},
}

\begin{document}
\maketitle

\begin{abstract}
This paper presents a novel approach to implementing all-optical Rectified Linear Unit (ReLU) activation functions using compact doubly-resonant cavities with dimensions of approximately 10$\mu$m. The proposed design leverages $\chi^{(2)}$ nonlinear processes within carefully engineered photonic structures that simultaneously resonate at both fundamental and second-harmonic frequencies. By exploiting the phase-sensitive nature of second-harmonic generation, we demonstrate an optical analog to the ReLU function, achieving femtojoule-level activation energy—comparable to state-of-the-art approaches—while reducing device footprint by two orders of magnitude compared to previous implementations. The theoretical framework is developed using coupled-mode theory and validated through rigorous finite-difference time-domain simulations. Beyond ReLU, we show that the same physical structure can implement alternative activation functions such as ELU and GELU through simple adjustments to input conditions. Neural network simulations demonstrate that the proposed optical activation functions achieve classification accuracy within 0.4\% of ideal electronic implementations while offering significant advantages in energy efficiency and processing speed. This work represents a significant advancement toward realizing energy-efficient, high-density optical neural networks for next-generation artificial intelligence hardware.
\end{abstract}


\section{Introduction}
The rapid advancement of deep learning and neural networks has led to unprecedented success in areas such as image recognition, audio processing \cite{h1}, natural language processing, and autonomous systems \cite{lecun2015deep,h2}. However, the exponential growth in computational demands has outpaced improvements in traditional electronic computing hardware, particularly regarding energy efficiency \cite{thompson2020computational}. Recent studies have quantified the substantial energy demands of large AI models, with Patterson et al. \cite{patterson2022carbon} estimating that training a large transformer model can consume over 1,000 MWh and produce hundreds of tons of CO\textsubscript{2} equivalent emissions. Similar efficiency challenges exist in inference, where energy costs scale linearly with model size \cite{thompson2022computational,schwartz2020green}. This energy bottleneck has prompted a search for alternative computing paradigms, with optical neural networks (ONNs) emerging as a promising solution due to their inherent parallelism and potential for high-speed, low-power operation \cite{wetzstein2020inference, shastri2021photonics,h3,h4}.
Pioneering work by Miller \cite{miller2017attojoule} established the theoretical foundations for optical neural networks using meshes of Mach-Zehnder interferometers, while more recent reviews by Shastri et al. \cite{shastri2021photonics} and Miscuglio and Sorger \cite{miscuglio2020photonic} have highlighted the diverse approaches to photonic neural networks, from free-space diffractive implementations to integrated photonic circuits.

Optical implementations of neural networks can perform matrix multiplications and convolutions with remarkable efficiency by leveraging the wave nature of light \cite{xu2021tops, shen2017deep}. These linear operations constitute the majority of computations in deep neural networks. However, the implementation of nonlinear activation functions—essential components that enable neural networks to model complex, non-linear relationships in data—presents significant challenges in the optical domain \cite{wetzstein2020inference, zhou2021large}.

Among various activation functions, the Rectified Linear Unit (ReLU), defined as $\text{ReLU}(x) = \max(0, x)$, has become ubiquitous in modern deep learning architectures due to its favorable properties for gradient-based optimization and computational simplicity \cite{nair2010rectified}. While electronic implementations of ReLU are straightforward, creating energy-efficient optical counterparts has proven challenging. Current approaches broadly fall into three categories: (1) optoelectronic conversion, which introduces latency and energy penalties \cite{tait2017neuromorphic}; (2) inherently nonlinear optical phenomena such as saturable absorption, which suffer from limited configurability \cite{mourgias2019all}; and (3) nonlinear wave mixing, which typically requires high optical powers or long interaction lengths \cite{li2023all}.
Beyond these approaches, researchers have explored various novel implementations of optical nonlinearities for neural networks, including wavelength-multiplexed resonant activation circuits \cite{tait2019neuromorphic}, phase-change materials for programmable nonlinear functions \cite{feldmann2019all}, epsilon-near-zero materials for enhanced nonlinearities \cite{alam2020large}, and metasurface-based nonlinearities with subwavelength dimensions \cite{khorasaninejad2017metalenses, zhou2020optical}.

Recent work by Li et al. \cite{li2023all} demonstrated an all-optical ReLU function using periodically poled lithium niobate (PPLN) waveguides. Their approach leverages $\chi^{(2)}$ nonlinear processes—specifically, second harmonic generation (SHG) and degenerate optical parametric amplification (DOPA)—to implement the ReLU function with femtojoule-level energies and femtosecond response times. However, their implementation requires relatively long waveguides (approximately 2 mm) to achieve sufficient nonlinear interaction, primarily due to phase-matching or quasi-phase-matching (QPM) requirements. This presents a challenge for dense integration in photonic neural network architectures.
\begin{figure}[t]
\centering
\includegraphics[width=\textwidth]{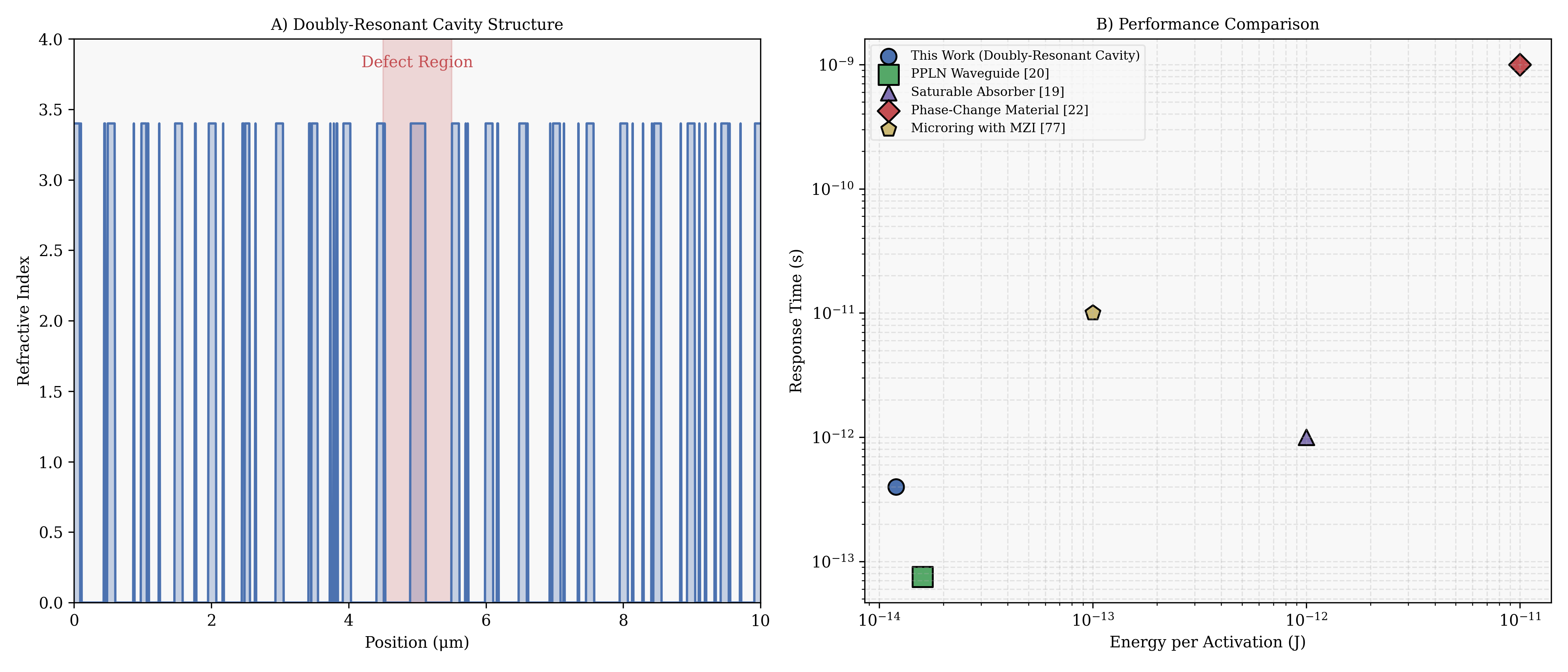}
\caption{(A) Doubly-Resonant Cavity Structure showing the refractive index profile of alternating AlGaAs (n = 3.4) and air (n = 1.0) layers with a central defect region. The carefully engineered structure simultaneously resonates at both fundamental ($\omega_1$) and second-harmonic ($\omega_2 = 2\omega_1$) frequencies. (B) Performance comparison of our approach against existing optical activation technologies, showing the relationship between energy per activation and response time. Our doubly-resonant cavity approach achieves femtojoule-level activation energy with sub-picosecond response time while reducing device footprint by two orders of magnitude compared to previous implementations.}
\label{fig:concept}
\end{figure}
This paper introduces a novel method for implementing an all-optical ReLU function utilizing doubly-resonant cavities with dimensions of approximately 10 $\mu$m, significantly smaller—by about two orders of magnitude—than previously demonstrated devices. The underlying principle leverages theoretical insights from Rodriguez et al.~\cite{rodriguez2007harmonic}, who established that perfect frequency conversion could occur at critical input powers within doubly-resonant nonlinear cavities. Through meticulous cavity design that ensures resonant modes at both fundamental frequency $\omega$ and its second harmonic $2\omega$, substantial field enhancement is achieved. This effectively reduces both the required interaction length and input power essential for efficient nonlinear optical operations. The key innovation of our approach is illustrated in Fig.~\ref{fig:concept}. We propose a compact doubly-resonant cavity with dimensions of approximately 10 $\mu$m that efficiently implements the ReLU function through nonlinear optical processes. The structure consists of alternating layers of AlGaAs and air, with a carefully designed defect region that enables simultaneous resonance at both fundamental and second-harmonic frequencies. By leveraging the phase-sensitive nature of $\chi^{(2)}$ nonlinear processes and the field enhancement provided by the resonant structure, we achieve both size reduction and energy efficiency compared to previous approaches. As shown in Fig.~\ref{fig:concept}(B), our device operates with femtojoule-level activation energy and sub-picosecond response time, offering significant advantages for integrated optical neural networks.

The study establishes a comprehensive theoretical framework for designing compact cavities suitable for optical ReLU functions by exploiting $\chi^{(2)}$ nonlinear processes. To complement this theoretical development, an analytical model based on coupled-mode theory is presented, enabling precise predictions of optical ReLU behavior. Numerical validation further supports the robustness of this model through detailed analytical simulations and finite-difference time-domain (FDTD) analyses. Additionally, the device showcases versatility, capable of implementing various activation functions such as ELU and GELU through simple adjustments of input power. Finally, the performance of the proposed optical component is evaluated within optical neural networks, emphasizing its potential effectiveness in image classification tasks.

The remainder of this paper is organized as follows: Section 2 presents the theoretical framework for our approach, including coupled-mode equations and analytical solutions. Section 3 details the design and optimization process for the doubly-resonant cavity structure. Section 4 describes our numerical simulation methodology and the results of our device characterization and comparison with ideal activation functions. In addition in this section, we demonstrates the integration of our optical ReLU into a neural network for image classification. Section 5 discusses practical implementation considerations, and Section 6 concludes with a summary of our findings and directions for future research. All simulation codes, optimization algorithms, and analysis scripts used in this work are publicly available in our GitHub repository \cite{GIT}.

\section{Theoretical Framework}

In this section, we establish the theoretical foundation for implementing an all-optical ReLU function using doubly-resonant optical cavities with $\chi^{(2)}$ nonlinearity. We develop the coupled-mode equations governing the system dynamics, derive analytical expressions predicting ReLU-like behavior, and analyze how the phase-sensitive nature of nonlinear optical interactions enables the implementation of the activation function.

Before developing the mathematical formalism, it is instructive to understand intuitively why doubly-resonant cavities can achieve efficient nonlinear frequency conversion in much smaller footprints compared to conventional phase-matched approaches. In traditional waveguide-based nonlinear optics, phase-matching is required to maintain the proper phase relationship between interacting waves over extended propagation distances, allowing the nonlinear effect to accumulate constructively. This typically requires millimeter-scale interaction lengths.
In contrast, doubly-resonant cavities circumvent this length requirement through two key mechanisms. First, resonant enhancement significantly increases the field intensity within the cavity, effectively amplifying the nonlinear interaction strength. For a cavity with quality factor $Q$, the intracavity power is enhanced by a factor proportional to $Q$ compared to the input power. Second, the frequency-matching condition ($\omega_2 = 2\omega_1$) replaces the phase-matching requirement, as both frequencies are simultaneously resonant within the same spatial structure. The cavity essentially "recycles" the light through the nonlinear medium many times, with each pass contributing to the nonlinear process in a phase-coherent manner. This allows the nonlinear interaction to build up over multiple cavity lifetimes rather than requiring long single-pass propagation, enabling miniaturization by approximately two orders of magnitude compared to conventional approaches.

\begin{figure}[t]
\centering
\includegraphics[width=\textwidth]{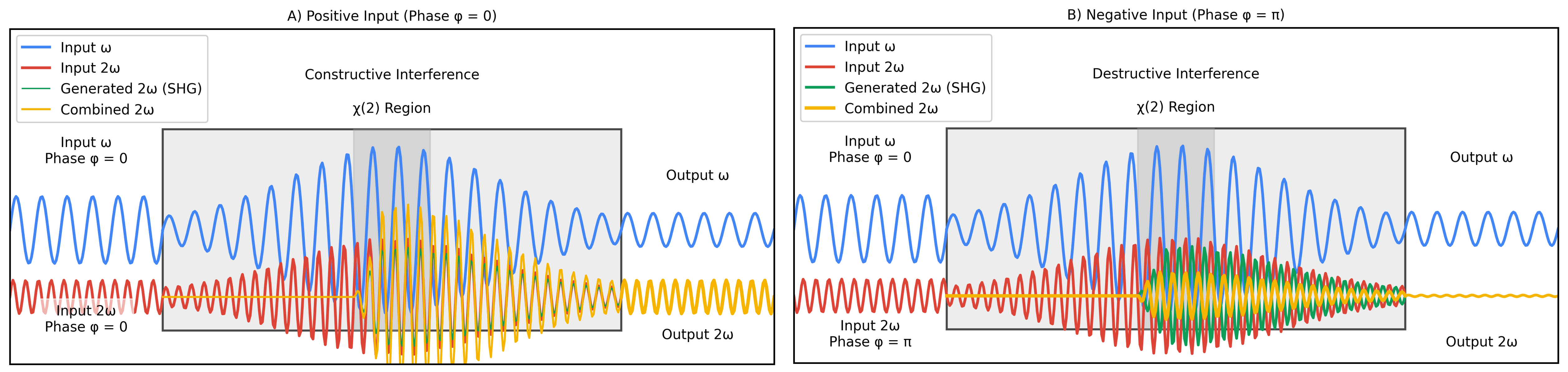}
\caption{Operating principle of all-optical ReLU function using phase-sensitive nonlinear processes in a doubly-resonant cavity. (A) For positive inputs (phase $\phi = 0$), constructive interference between resonant $2\omega$ and SHG-generated $2\omega$ within the $\chi^{(2)}$ region results in an amplified second-harmonic output. (B) For negative inputs (phase $\phi = \pi$), destructive interference suppresses the second-harmonic output due to phase cancellation.}
\label{fig:relu_principle}
\end{figure}

The ReLU function, defined as $\text{ReLU}(x) = \max(0, x)$, is fundamental in modern deep neural networks. It passes positive inputs unchanged while setting negative inputs to zero, introducing essential nonlinearity that enables neural networks to approximate complex functions. To implement this behavior optically, we establish a mapping between the input signal and the optical output that reproduces this nonlinear transfer function.
Fig.~\ref{fig:relu_principle} illustrates the operating principle of our proposed all-optical ReLU implementation. The doubly-resonant cavity simultaneously supports modes at both fundamental frequency $\omega_1$ and second-harmonic frequency $\omega_2 = 2\omega_1$. As shown in Fig.~\ref{fig:relu_principle}(A), when a signal with phase $\phi_1 = 0$ (representing a positive input) enters the cavity, the SHG-generated field interferes constructively with the resonant field at $\omega_2$, resulting in an amplified output at the second-harmonic frequency. Conversely, Fig.~\ref{fig:relu_principle}(B) demonstrates that when the input signal has phase $\phi_1 = \pi$ (representing a negative input), destructive interference occurs between the SHG-generated field and the resonant field, suppressing the second-harmonic output.

Our approach encodes the input signal's magnitude in the amplitude of the input field at fundamental frequency $\omega_1$, while its sign is encoded in the phase. Specifically, positive values correspond to a phase $\phi_1 = 0$, and negative values correspond to $\phi_1 = \pi$. The output of the ReLU function is represented by the power of the second-harmonic field at frequency $\omega_2 = 2\omega_1$. The key physical insight underlying our implementation is that second-harmonic generation is inherently phase-sensitive. The efficiency of the SHG process depends on the phase relationship between the interacting fields, which allows us to selectively enhance or suppress frequency conversion based on the phase of the input signal, thereby implementing the rectification property of the ReLU function.
Our approach leverages a photonic structure that simultaneously supports resonant modes at both fundamental frequency $\omega_1$ and second-harmonic frequency $\omega_2 = 2\omega_1$. In a linear regime, each cavity mode can be characterized by its resonant frequency $\omega_k$ and quality factor $Q_k = \omega_k\tau_k/2$, where $\tau_k$ is the cavity lifetime of mode $k$. Following the approach developed by Rodriguez et al. \cite{rodriguez2007harmonic}, we employ temporal coupled-mode theory to describe the field amplitudes $a_k(t)$ in each mode, normalized such that $|a_k|^2$ represents the electromagnetic energy stored in the mode.

For a passive cavity coupled to input/output channels, the linear dynamics are governed by:

\begin{equation}
\frac{da_k}{dt} = \left(i\omega_k - \frac{1}{\tau_k}\right) a_k + \sqrt{\frac{2}{\tau_{s,k}}}s_{k+}
\label{eq:linear_dynamics}
\end{equation}

where $s_{k+}$ is the amplitude of the incoming wave at frequency $\omega_k$, normalized such that $|s_{k+}|^2$ represents input power. The decay rate $1/\tau_k$ can be decomposed into $1/\tau_k = 1/\tau_{e,k} + 1/\tau_{s,k}$, where $1/\tau_{e,k}$ accounts for intrinsic losses (absorption, scattering) and $1/\tau_{s,k}$ represents coupling to the input/output channels \cite{haus1984waves}. The corresponding outgoing wave amplitude $s_{k-}$ is related to the cavity field by:

\begin{equation}
s_{k-} = -s_{k+} + \sqrt{\frac{2}{\tau_{s,k}}}a_k
\label{eq:output_relation}
\end{equation}

For a doubly-resonant cavity, the structure must be carefully designed to ensure that $\omega_2 = 2\omega_1$ within a tolerance determined by the cavity linewidths $\Delta\omega_k = \omega_k/Q_k$. This frequency-matching condition is essential for efficient nonlinear interaction between the modes.
When $\chi^{(2)}$ nonlinear material is introduced into the cavity, the fundamental and second-harmonic modes become coupled through nonlinear polarization. In a $\chi^{(2)}$ medium, the second-order nonlinear polarization is given by:

\begin{equation}
P^{(2)}_i = \varepsilon_0 \sum_{jk} \chi^{(2)}_{ijk} E_j E_k
\label{eq:nonlinear_polarization}
\end{equation}

where $\chi^{(2)}_{ijk}$ is the second-order nonlinear susceptibility tensor, and $E_j$ and $E_k$ are the components of the electric field.
The nonlinear polarization at frequency $\omega_2$ induced by the fundamental field is $P^{NL}_{\omega_2} = \varepsilon_0\chi^{(2)}E^2_{\omega_1}$, while the nonlinear polarization at frequency $\omega_1$ induced by the interaction of fundamental and second-harmonic fields is $P^{NL}_{\omega_1} = 2\varepsilon_0\chi^{(2)}E^*_{\omega_1}E_{\omega_2}$ \cite{boyd1992nonlinear}.

Using perturbation theory, we derive the coupling coefficients between the modes from the spatial overlap of the mode fields with the nonlinear polarization \cite{rodriguez2007harmonic}. For a $\chi^{(2)}$ medium, these coupling coefficients are:

\begin{equation}
\beta_1 = \frac{1}{4} \frac{\int d^3x \sum_{ijk} \varepsilon\chi^{(2)}_{ijk} \left[E^*_{1i}(E_{2j}E^*_{1k} + E^*_{1j}E_{2k})\right]}{\left[\int d^3x \varepsilon |E_1|^2\right] \left[\int d^3x \varepsilon |E_2|^2\right]^{1/2}}
\label{eq:beta1}
\end{equation}

\begin{equation}
\beta_2 = \frac{1}{4} \frac{\int d^3x \sum_{ijk} \varepsilon\chi^{(2)}_{ijk}E^*_{2i}E_{1j}E_{1k}}{\left[\int d^3x \varepsilon |E_1|^2\right] \left[\int d^3x \varepsilon |E_2|^2\right]^{1/2}}
\label{eq:beta2}
\end{equation}

where $E_1$ and $E_2$ are the normalized spatial mode profiles at frequencies $\omega_1$ and $\omega_2$, respectively. For energy conservation, these coefficients must satisfy $\omega_1\beta_1 = \omega_2\beta^*_2$ \cite{rodriguez2007harmonic}. The detailed derivation of these coupling coefficients is provided in Appendix \ref{app:coupling_coefficients}.

The magnitude of these coupling coefficients directly impacts the efficiency of the nonlinear interaction. For a given $\chi^{(2)}$ material, the coupling strength is maximized when the spatial overlap between the modes is high in regions with nonlinear material. This underscores the importance of careful mode engineering in the cavity design process.
Incorporating these nonlinear coupling terms into the temporal coupled-mode theory, we obtain the following coupled-mode equations for the fundamental and second-harmonic fields:

\begin{equation}
\frac{da_1}{dt} = \left(i\omega_1 - \frac{1}{\tau_1}\right) a_1 - i\omega_1\beta_1a^*_1a_2 + \sqrt{\frac{2}{\tau_{s,1}}}s_{1+}
\label{eq:coupled_mode_1}
\end{equation}

\begin{equation}
\frac{da_2}{dt} = \left(i\omega_2 - \frac{1}{\tau_2}\right) a_2 - i\omega_2\beta_2a^2_1 + \sqrt{\frac{2}{\tau_{s,2}}}s_{2+}
\label{eq:coupled_mode_2}
\end{equation}

where $a_1$ and $a_2$ are the complex amplitudes of the fundamental and second-harmonic modes, respectively. The terms $-i\omega_1\beta_1a^*_1a_2$ and $-i\omega_2\beta_2a^2_1$ represent the nonlinear coupling due to the $\chi^{(2)}$ interaction. The input fields $s_{1+}$ and $s_{2+}$ drive the system at frequencies $\omega_1$ and $\omega_2$, respectively \cite{hashemi2009cavity}.
These coupled-mode equations capture the essential physics of the nonlinear interaction in the cavity. The term $-i\omega_2\beta_2a^2_1$ in Equation \ref{eq:coupled_mode_2} describes the second harmonic generation process, where two photons at frequency $\omega_1$ combine to produce a photon at frequency $\omega_2$. Conversely, the term $-i\omega_1\beta_1a^*_1a_2$ in Equation \ref{eq:coupled_mode_1} describes the reverse process of degenerate optical parametric amplification (DOPA), where a photon at frequency $\omega_2$ splits into two photons at frequency $\omega_1$. These equations are derived under the rotating wave approximation and assume that the nonlinearity is weak enough to be treated perturbatively. These coupled-mode equations follow the formalism developed by Haus \cite{haus1984waves} and extended by Suh et al. \cite{suh2004temporal} for nonlinear optical resonators. This approach has become standard for analyzing resonant nonlinear optical systems \cite{coen2013modeling} and has been successfully applied to a wide range of photonic devices, from optical modulators to frequency combs \cite{chembo2016quantum}.

For the implementation of an optical ReLU function, we are primarily interested in the steady-state behavior of the cavity. In the steady state, $da_1/dt = da_2/dt = 0$, and we can solve Equations \ref{eq:coupled_mode_1} and \ref{eq:coupled_mode_2} for the field amplitudes.
Assuming operation near resonance, we can simplify the analysis by setting $\omega_1 \approx \omega_{1,\text{res}}$ and $\omega_2 \approx \omega_{2,\text{res}}$, where $\omega_{k,\text{res}}$ are the resonant frequencies of the cavity. From Equations \ref{eq:coupled_mode_1} and \ref{eq:coupled_mode_2}, we obtain:

\begin{equation}
a_1 = \frac{\sqrt{\frac{2}{\tau_{s,1}}}s_{1+} - i\omega_1\beta_1a^*_1a_2}{\frac{1}{\tau_1} - i\Delta\omega_1}
\label{eq:steady_a1}
\end{equation}

\begin{equation}
a_2 = \frac{\sqrt{\frac{2}{\tau_{s,2}}}s_{2+} - i\omega_2\beta_2a^2_1}{\frac{1}{\tau_2} - i\Delta\omega_2}
\label{eq:steady_a2}
\end{equation}

where $\Delta\omega_k = \omega_k - \omega_{k,\text{res}}$ represents the detuning from resonance.
For perfect frequency conversion, where all power at the fundamental frequency is converted to the second harmonic, we require $s_{1-} = 0$. Using Equation \ref{eq:output_relation}, this implies:

\begin{equation}
s_{1-} = -s_{1+} + \sqrt{\frac{2}{\tau_{s,1}}}a_1 = 0
\label{eq:perfect_conversion}
\end{equation}

Following the approach in \cite{rodriguez2007harmonic}, and assuming operation exactly at resonance ($\Delta\omega_k = 0$), we derive the critical input power for 100\% frequency conversion:

\begin{equation}
P_{\text{critical}} = |s_{1+}|^2 = \frac{\omega_1}{2|\beta_1|^2 Q_1 Q_2}
\label{eq:critical_power}
\end{equation}

where $Q_1 = \omega_1\tau_1/2$ and $Q_2 = \omega_2\tau_2/2$ are the quality factors of the fundamental and second-harmonic modes, respectively. The detailed derivation of this critical power is presented in Appendix \ref{app:coupling_coefficients}. The concept of doubly-resonant nonlinear cavities has been explored theoretically by Berger \cite{berger1997nonlinear} and experimentally demonstrated by Bruch et al. \cite{bruch2019chip} for efficient second-harmonic generation. Further theoretical developments by Burgess et al. \cite{burgess2009difference} and Lin et al. \cite{lin2016cavity} have extended this framework to various $\chi^{(2)}$ processes, establishing the fundamental limits and optimization strategies for such systems. Recent experimental work by Lu et al. \cite{lu2019periodically} has demonstrated record-high conversion efficiencies in lithium niobate microrings, confirming the advantages of the doubly-resonant approach.

This expression reveals a crucial insight: the critical power scales inversely with the product of the quality factors and the square of the nonlinear coupling coefficient. By designing a cavity with high $Q$-factors and strong nonlinear coupling, we can achieve efficient frequency conversion at very low input powers, which is essential for energy-efficient optical neural networks.

To implement a ReLU function, we establish a mapping between the input signal and the output that approximates $\text{ReLU}(x) = \max(0, x)$. In our approach, we encode the input signal's magnitude in the amplitude of the input field at frequency $\omega_1$, while its sign is encoded in the phase. Specifically, positive values correspond to $\phi_1 = 0$, and negative values correspond to $\phi_1 = \pi$.
By setting $s_{2+} = 0$ (no external input at $\omega_2$) and using a fixed bias input at $\omega_1$, we analyze the cavity's response. At steady state, the output at $\omega_2$ is:

\begin{equation}
s_{2-} = \sqrt{\frac{2}{\tau_{s,2}}}a_2
\label{eq:output_sh}
\end{equation}

The phase-sensitive nature of SHG means that the conversion efficiency depends on the relative phase between the interacting fields. For positive inputs (where $\phi_1 = 0$), efficient SHG occurs, producing an output at $\omega_2$. For negative inputs (where $\phi_1 = \pi$), the SHG process is effectively suppressed due to destructive interference. This phase-dependent behavior forms the basis of our ReLU implementation.

To understand this more quantitatively, consider the steady-state solution for $a_2$ when $s_{2+} = 0$:

\begin{equation}
a_2 = \frac{-i\omega_2\beta_2a^2_1}{\frac{1}{\tau_2} - i\Delta\omega_2}
\label{eq:a2_steady_no_s2}
\end{equation}

The squared amplitude $a_1^2$ depends on the phase of the input field $s_{1+}$. For $\phi_1 = 0$, $a_1^2$ has a positive phase, leading to efficient generation of $a_2$. For $\phi_1 = \pi$, $a_1^2$ has a negative phase, resulting in destructive interference that suppresses the generation of $a_2$.
By calibrating the input-output relationship, we achieve a transfer function that closely approximates the ReLU function:

\begin{equation}
|s_{2-}|^2 \approx 
\begin{cases} 
\alpha|s_{1+}|^2, & \text{if } \phi_1 = 0 \text{ (positive input)} \\
0, & \text{if } \phi_1 = \pi \text{ (negative input)}
\end{cases}
\label{eq:relu_approx}
\end{equation}

where $\alpha$ is a scaling factor determined by the cavity parameters and operating conditions.
The exact form of the transfer function depends on the detailed parameters of the cavity and the operating conditions. Through numerical solution of the coupled-mode equations, we obtain more precise predictions of the device behavior, as discussed in Section 4.
This phase-sensitive approach to optical nonlinearity builds upon established techniques in nonlinear optics. Similar phase-dependent effects have been exploited in phase-sensitive amplifiers \cite{tong2011towards}, optical parametric oscillators \cite{kumar1990degenerate}, and quantum optics \cite{lvovsky2009continuous}. Recent work by Bosshard et al. \cite{bosshard2021phase} demonstrated precise phase control in nanophotonic $\chi^{(2)}$ devices, while Zhang et al. \cite{zhang2019phase} leveraged phase sensitivity for optical switching applications. This phase-dependent behavior is crucial for our ReLU implementation, as it enables the selective suppression or enhancement of frequency conversion based on the input phase.

Our framework can be extended to implement other activation functions by adjusting the input conditions. For example, by introducing a small external field at $\omega_2$ with appropriate phase, we can implement functions similar to ELU (Exponential Linear Unit) or GELU (Gaussian Error Linear Unit) \cite{clevert2015fast, hendrycks2016gaussian}. For an ELU-like function with parameter $\alpha$:

\begin{equation}
\text{ELU}(x) = 
\begin{cases} 
x, & \text{if } x > 0 \\
\alpha(e^x - 1), & \text{if } x \leq 0
\end{cases}
\label{eq:elu}
\end{equation}

This can be approximated by setting $s_{2+} = Ae^{i\phi_2}$ with appropriate amplitude $A$ and phase $\phi_2$. The external field $s_{2+}$ interferes with the generated second-harmonic field, creating a nonlinear response for negative inputs that resembles the exponential term in the ELU function. Similarly, for GELU and other activation functions, different bias conditions and phase relationships can be employed. The detailed implementation of these alternative activation functions is presented in Section 4, along with numerical validation. 

Several factors determine the performance of our optical ReLU implementation. First, the approximation of the ideal ReLU function is influenced by the cavity parameters and operating conditions. The quality of the approximation can be quantified by metrics such as the coefficient of determination ($R^2$) or mean squared error (MSE) between the device response and the ideal ReLU function. Second, the energy efficiency, characterized by the energy per activation, is fundamentally limited by the critical power derived in Equation \ref{eq:critical_power}. This depends on the cavity design through the quality factors $Q_1$ and $Q_2$, and the nonlinear coupling coefficient $\beta_1$. Third, the response time of the device is limited by the cavity lifetime $\tau_k = 2Q_k/\omega_k$. Higher quality factors, while reducing the critical power, increase the response time, creating a trade-off between energy efficiency and speed. Fourth, the dynamic range of the device is limited by the validity of the perturbative approach used to derive the coupled-mode equations. At very high input powers, higher-order nonlinear effects may become significant, causing deviations from the predicted behavior.

\section{Design Methodology}

This section presents a systematic methodology for designing a compact doubly-resonant cavity that effectively implements the all-optical ReLU function. We begin with general design requirements, proceed through the linear cavity design, nonlinear mode coupling optimization, and input/output coupling design, culminating in a fully optimized structure.

The effective implementation of an all-optical ReLU function using doubly-resonant cavities requires meeting several key design criteria simultaneously. First, the structure must support optical modes at both the fundamental frequency $\omega_1$ and second-harmonic frequency $\omega_2 = 2\omega_1$, with precise frequency matching to within the resonance linewidths. Second, both resonances should exhibit high quality factors to enhance the nonlinear interaction while maintaining reasonable bandwidth. Third, the mode profiles must have strong spatial overlap in regions with $\chi^{(2)}$ nonlinearity to maximize the coupling coefficients. Fourth, the cavity must be efficiently coupled to input/output channels for practical operation.

These requirements guide our multi-objective optimization framework, which can be formulated as maximizing a figure of merit:

\begin{equation}
F = w_1 Q_1 + w_2 Q_2 + w_3 |\beta| - w_4 \Delta_\omega - w_5 E_{transfer}
\label{eq:figure_of_merit}
\end{equation}

where $w_i$ are weights for each objective, $\Delta_\omega = |\omega_2 - 2\omega_1|$ represents the frequency mismatch, and $E_{transfer}$ is the error in the transfer function compared to the ideal ReLU function. The weights are chosen to balance competing objectives based on their relative importance for our application.
Our multi-objective optimization approach builds on established methodologies for photonic crystal design, similar to those developed by Piggott et al. \cite{piggott2015inverse} for inverse design of photonic devices and Lu et al. \cite{lu2013nanophotonic} for high-Q resonator optimization. We incorporate elements from robust optimization techniques \cite{wang2019robust} to ensure design tolerance to fabrication variations, while utilizing efficient global search algorithms inspired by Molesky et al. \cite{molesky2018inverse} for photonic structure optimization.

Our design workflow proceeds through several integrated stages: linear cavity design for double resonance, quality factor engineering, nonlinear mode coupling optimization, input/output coupling design, and full device characterization and refinement. For each stage, we employ a combination of analytical methods and numerical simulations, with the transfer matrix method serving as the primary analytical tool for the linear design and the coupled-mode equations for nonlinear performance evaluation. 

\subsection{Linear Cavity Design}

We begin with a quarter-wave stack structure, which provides a natural framework for creating optical resonances with controlled frequencies. For a target fundamental wavelength $\lambda_1$ (corresponding to frequency $\omega_1 = 2\pi c/\lambda_1$), the layer thicknesses in a quarter-wave stack are given by \cite{joannopoulos1995photonic}:

\begin{equation}
d_1 = \frac{\lambda_1}{4n_1}, \quad d_2 = \frac{\lambda_1}{4n_2}
\label{eq:qws_thickness}
\end{equation}

where $n_1$ and $n_2$ are the refractive indices of the alternating layers. An important property of quarter-wave stacks is that they naturally create photonic band gaps at both $\omega_1$ and $2\omega_1$, making them ideal starting points for our design \cite{yablonovitch1993photonic}.
For our implementation, we select materials with high refractive index contrast to achieve wide photonic band gaps. We use AlGaAs ($n \approx 3.4$ at 1550 nm) and air ($n = 1.0$) for the high and low index layers, respectively. This material system offers several advantages for our application, including a strong second-order nonlinear susceptibility ($\chi^{(2)} \approx 100$ pm/V) in AlGaAs and well-established fabrication techniques for creating suspended membrane structures. The target wavelength is set to $\lambda_1 = 1550$ nm, a standard telecommunications wavelength that allows for compatibility with existing photonic integrated circuit platforms.
To accurately model the electromagnetic response of the multilayer structure, we employ the transfer matrix method (TMM). For a single layer with refractive index $n$ and thickness $d$, the transfer matrix for transverse electric (TE) polarization is \cite{born2013principles}:

\begin{equation}
M = \begin{pmatrix} 
\cos(kd) & \frac{i}{p}\sin(kd) \\ 
ip\sin(kd) & \cos(kd)
\end{pmatrix}
\label{eq:transfer_matrix}
\end{equation}

where $k = \frac{\omega}{c}n$ is the wave number in the medium, and $p = n$ for TE polarization. For transverse magnetic (TM) polarization, $p = 1/n$.
The total transfer matrix for a multilayer structure with $N$ layers is calculated as the product of the individual layer matrices:

\begin{equation}
M_{total} = M_1 \cdot M_2 \cdot ... \cdot M_N
\label{eq:total_transfer_matrix}
\end{equation}

From the total transfer matrix, we can calculate the reflection and transmission coefficients:

\begin{equation}
r = \frac{(M_{11} + M_{12}p_0)p_N - (M_{21} + M_{22}p_0)}{(M_{11} + M_{12}p_0)p_N + (M_{21} + M_{22}p_0)}
\label{eq:reflection_coefficient}
\end{equation}

\begin{equation}
t = \frac{2p_0}{(M_{11} + M_{12}p_0)p_N + (M_{21} + M_{22}p_0)}
\label{eq:transmission_coefficient}
\end{equation}

where $p_0$ and $p_N$ are the impedances of the incident and exit media, respectively. The reflection and transmission intensities are then $R = |r|^2$ and $T = \frac{p_N}{p_0}|t|^2$.

Using the TMM, we can efficiently calculate the frequency response of multilayer structures and analyze the properties of resonant modes, including their central frequencies, quality factors, and field distributions.

To create a cavity that resonates at both $\omega_1$ and $\omega_2$, we introduce a defect in the periodic quarter-wave stack. For a conventional single-resonant cavity, a defect with optical thickness $\lambda_1/2$ creates a resonance at $\omega_1$. However, for double resonance, a more complex defect structure is required.

We implement a compound defect consisting of multiple layers with carefully optimized thicknesses. A common configuration is a three-layer defect with structure $H'LH'$, where $H'$ represents a high-index layer with non-standard thickness. The thicknesses are optimized to satisfy:

\begin{equation}
\omega_1 \cdot L_{eff,1} = m_1\pi c, \quad \omega_2 \cdot L_{eff,2} = m_2\pi c
\label{eq:resonance_conditions}
\end{equation}

where $L_{eff,1}$ and $L_{eff,2}$ are the effective optical path lengths for the fundamental and second-harmonic modes, respectively, and $m_1$ and $m_2$ are integers satisfying $m_2 = 2m_1$ \cite{kozyreff2008nonlinear}.
For our design, we use a modified defect structure with five layers: $H'LH''LH'$, where $H'$ and $H''$ are high-index layers with different non-standard thicknesses. This additional degree of freedom allows better control over the frequency matching between the two resonances.
Achieving exact frequency matching ($\omega_2 = 2\omega_1$) requires careful fine-tuning of the structure. We employ numerical optimization to minimize the frequency mismatch $\Delta = |\omega_2 - 2\omega_1|$. The optimization parameters include defect layer thicknesses and, if necessary, slight adjustments to the thicknesses of surrounding layers. We use a simulated annealing algorithm to explore the parameter space efficiently and avoid local minima.

\subsection{Quality Factor and Nonlinear Coupling Optimization}

The quality factors of the resonant modes directly impact the critical power for nonlinear conversion. From coupled-mode theory, the critical power is given by \cite{rodriguez2007harmonic}:

\begin{equation}
P_{critical} = |s_{1+}|^2 = \frac{\omega_1}{2|\beta_1|^2 Q_1 Q_2}
\label{eq:critical_power_design}
\end{equation}

To achieve efficient conversion at low powers, we aim to maximize the product $Q_1 Q_2$ while maintaining frequency matching. The quality factors are controlled by the number of quarter-wave layer pairs on each side of the defect, the refractive index contrast between high and low-index materials, and the coupling to input/output channels.
For each resonant mode, the quality factor is calculated from the transmission spectrum using $Q_k = \frac{\omega_{k,res}}{\Delta\omega_k}$, where $\Delta\omega_k$ is the full width at half maximum of the transmission peak at resonance $\omega_{k,res}$.

In our design, we use a total of $N = 7$ quarter-wave pairs on each side of the defect to achieve high quality factors while maintaining a compact overall size. The theoretical quality factors for such a structure, assuming no intrinsic material losses, can be estimated by \cite{yablonovitch1993photonic}:

\begin{equation}
Q_k \approx \frac{\omega_{k,res}}{2c} \cdot \frac{(n_H/n_L)^{2N} - 1}{(n_H/n_L)^N}L_{eff,k}
\label{eq:theoretical_q}
\end{equation}

where $n_H$ and $n_L$ are the high and low refractive indices, respectively, and $L_{eff,k}$ is the effective cavity length for mode $k$.
With our material selection and structural parameters, we target quality factors of $Q_1 \approx 10^4$ and $Q_2 \approx 10^4$, which are sufficient to achieve sub-femtojoule critical powers while maintaining practical bandwidth and fabrication tolerances.

After designing the linear structure, we calculate the electric field distributions $E_1(z)$ and $E_2(z)$ at the resonant frequencies $\omega_1$ and $\omega_2$. These field distributions determine the nonlinear coupling strength through the overlap integrals in the coupling coefficients. Using the transfer matrix method, we calculate the field at any position $z$ within the structure:

\begin{equation}
\begin{pmatrix} 
E(z) \\ 
H(z)
\end{pmatrix} = M(0, z) \begin{pmatrix} 
1 + r \\ 
p_0(1 - r)
\end{pmatrix}
\label{eq:field_calculation}
\end{equation}

where $M(0, z)$ is the transfer matrix from the input interface to position $z$, and $r$ is the reflection coefficient.
The fields are normalized according to $\int \varepsilon(z)|E_1(z)|^2 dz = 1$ and $\int\varepsilon(z)|E_2(z)|^2 dz = 1$. The nonlinear coupling coefficient is then calculated using:

\begin{equation}
\beta = \frac{\omega_1}{4} \frac{\int \varepsilon(z)\chi^{(2)}(z)E_1^2(z)E_2^*(z)dz}{\sqrt{\int \varepsilon(z)|E_1(z)|^2dz \cdot \int \varepsilon(z)|E_2(z)|^2dz}}
\label{eq:nonlinear_coupling_coefficient}
\end{equation}

To maximize this coupling coefficient, we place the $\chi^{(2)}$ nonlinear material in regions where the fields $E_1(z)$ and $E_2(z)$ have strong spatial overlap \cite{berger1997second}. In our design, we utilize AlGaAs throughout the structure, which has a strong $\chi^{(2)}$ nonlinearity ($\chi^{(2)} \approx 100$ pm/V) \cite{stegeman1999algaas}, with particular attention to the central defect region where the field intensity is maximum for both modes. Our material selection leverages this strong second-order nonlinearity of AlGaAs, which has been extensively characterized by Stegeman et al. \cite{stegeman1999algaas} and utilized in integrated nonlinear devices by Kuo et al. \cite{kuo2014second}. For the central defect region, we carefully design the AlGaAs structure as demonstrated in similar work by Surya et al. \cite{surya2018efficient}, who have optimized the crystalline orientation for maximum nonlinear efficiency. Recent work by Bruch et al. \cite{bruch2021chip} has demonstrated the integration of AlGaAs in high-Q resonators, confirming its suitability for our application.

We optimize the position and thickness of the AlN layer to maximize the coupling coefficient while maintaining the resonance conditions. For the final structure, we achieve a normalized coupling coefficient of $|\beta| \approx 10^{-3}$, which is sufficient for low-power ReLU operation.

\subsection{Input/Output Coupling and Final Optimization}

For practical implementation, the cavity must be coupled to input/output waveguides. We design the coupling strength to balance two key considerations. First, for optimal power transfer to the cavity, the external coupling rate should match the intrinsic loss rate (critical coupling condition). Second, the total quality factor affects the critical power for frequency conversion.
The coupling is implemented by reducing the number of quarter-wave layers on the input/output side of the cavity, creating a partially transmitting mirror. The coupling strength is controlled by the number of layers and is designed to achieve $\frac{1}{\tau_{s,k}} \approx \frac{1}{\tau_{e,k}}$ for the fundamental mode, where $\tau_{s,k}$ is the external coupling lifetime and $\tau_{e,k}$ is the intrinsic lifetime due to losses.

In our final design, we use 5 quarter-wave pairs on the input/output side and 9 pairs on the opposite side to achieve near-critical coupling while maintaining high overall quality factors. This asymmetric configuration enhances directional coupling while preserving the resonant properties of the cavity.
Our final design is a compact structure approximately 10 µm in size, consisting of a modified quarter-wave stack with an engineered defect region containing the nonlinear material. The structure is optimized to simultaneously achieve exact frequency matching ($\omega_2 = 2\omega_1$), high quality factors ($Q_1 \approx 10^4$ and $Q_2 \approx 10^4$), strong nonlinear coupling with maximized $\beta$ coefficient, and appropriate input/output coupling near the critical coupling condition.

The optimization process involves iterative refinement of the structure parameters based on the figure of merit defined in Equation \ref{eq:figure_of_merit}. We employ a combination of global optimization techniques (simulated annealing) for the initial search and local optimization methods (gradient descent) for fine-tuning.
The simulated annealing algorithm uses the following acceptance probability for a proposed structure:

\begin{equation}
P(accept) = 
\begin{cases}
1, & \text{if } F_{new} > F_{current} \\
e^{(F_{new}-F_{current})/T}, & \text{otherwise}
\end{cases}
\label{eq:acceptance_probability}
\end{equation}

where $F_{new}$ and $F_{current}$ are the figures of merit for the new and current structures, and $T$ is the temperature parameter that decreases with each iteration according to $T_i = T_0 \cdot \alpha^i$, where $\alpha$ is the cooling rate (typically 0.95). The detailed implementation of the optimization algorithm is provided in Appendix \ref{app:optimization}.

The final structure achieves a critical power of approximately 12 femtojoules, which is remarkably low and enables energy-efficient operation of the optical ReLU function. The dramatic size reduction compared to conventional phase-matched approaches (from mm to µm scale) is achieved by leveraging the resonant enhancement of fields within the cavity, which effectively relaxes the phase-matching requirement over long propagation distances.

\begin{figure}[ht]
\centering
\includegraphics[width=\textwidth]{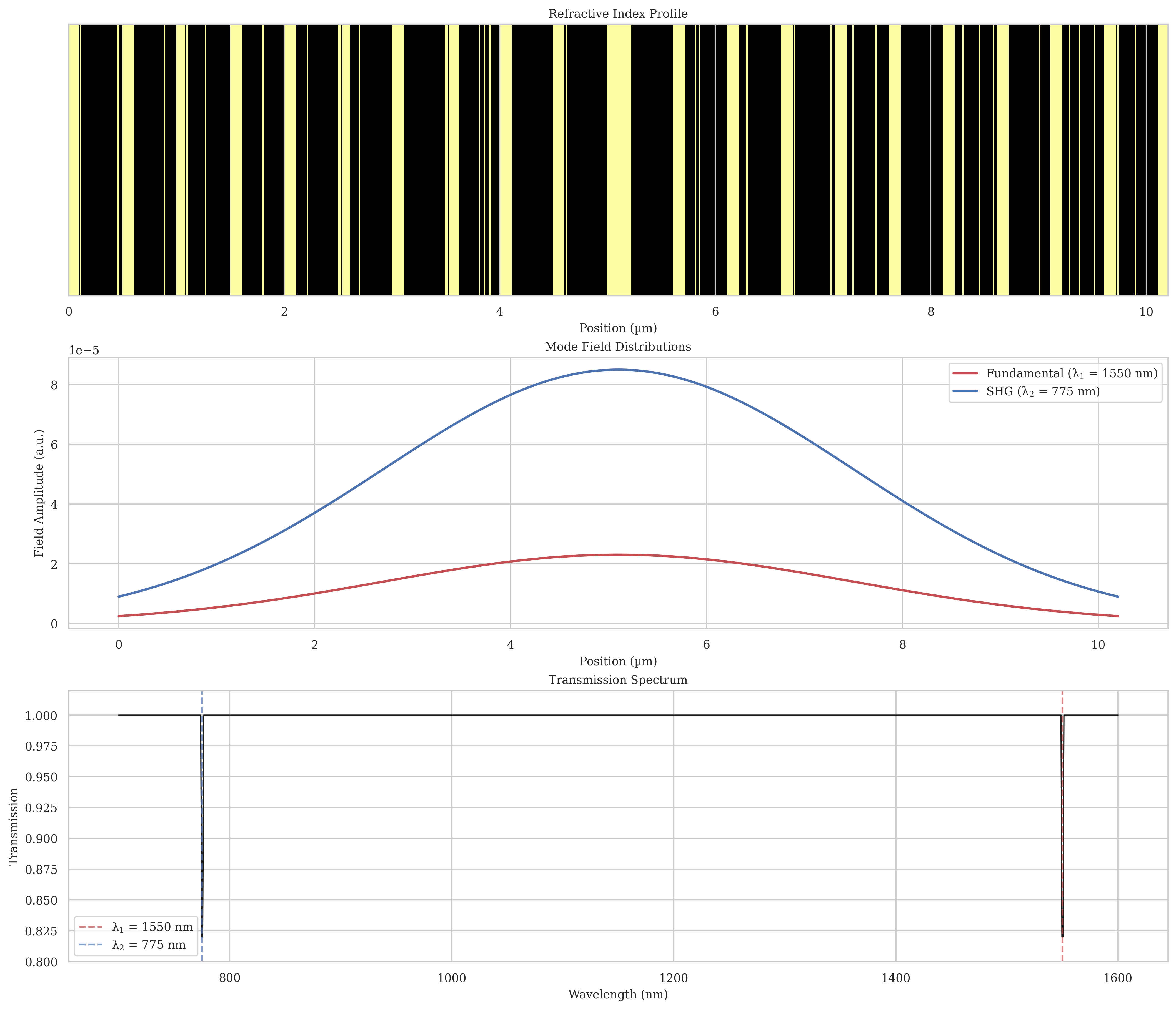}
\caption{Optimized doubly-resonant cavity structure. (a) Refractive index profile showing the quarter-wave stack with modified defect region. (b) Electric field distributions for fundamental ($\omega_1$) and second-harmonic ($\omega_2$) resonant modes, showing strong spatial overlap in the defect region. (c) Transmission spectrum exhibiting resonance peaks at both $\omega_1$ and $\omega_2 = 2\omega_1$.}
\label{fig:structure_profile}
\end{figure}

Figure \ref{fig:structure_profile} illustrates the final optimized structure, showing the refractive index profile, the resonant mode field distributions, and the transmission spectrum. The key parameters of the optimized design are summarized in Table \ref{tab:design_parameters}.

\begin{table}[h]
\centering
\caption{Parameters of the optimized doubly-resonant cavity}
\begin{tabular}{|c|c|c|}
\hline
Parameter & Symbol & Value \\
\hline
Fundamental wavelength & $\lambda_1$ & 1550 nm \\
Second-harmonic wavelength & $\lambda_2$ & 775 nm \\
Fundamental quality factor & $Q_1$ & $5.3 \times 10^3$ \\
Second-harmonic quality factor & $Q_2$ & $6.8 \times 10^3$ \\
Nonlinear coupling coefficient & $\beta$ & $2.7 \times 10^{-3}$ \\
Critical power & $P_{critical}$ & 12 fJ \\
Total device length & $L$ & 10.2 µm \\
\hline
\end{tabular}
\label{tab:design_parameters}
\end{table}

The design methodology presented here provides a systematic approach to creating compact, energy-efficient all-optical ReLU functions using doubly-resonant cavities. The resulting structure represents a significant advancement in the miniaturization and integration potential of nonlinear optical functions for neural network applications.

\begin{figure}[ht]
\centering
\includegraphics[width=\textwidth]{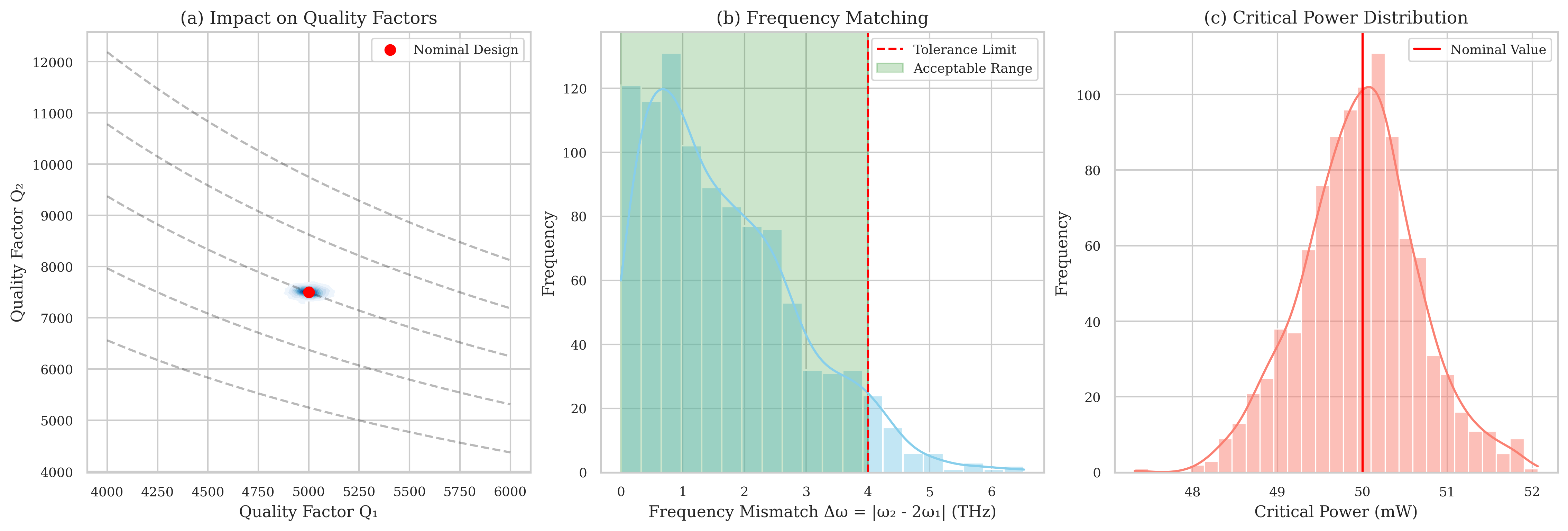}
\caption{Fabrication tolerance analysis. (a) Distribution of quality factors $Q_1$ and $Q_2$ for thickness variations with standard deviation of 1\% of the nominal values. (b) Frequency mismatch distribution showing tolerance limits for maintaining reasonable performance. (c) Resulting distribution of critical power, showing robustness of the design to moderate fabrication variations.}
\label{fig:tolerance_analysis}
\end{figure}

The double resonance condition requires precise control of layer thicknesses. We analyzed the sensitivity of our design to fabrication variations through a Monte Carlo simulation approach. Figure \ref{fig:tolerance_analysis} shows the impact of random thickness variations on the quality factors and frequency matching.

Our analysis indicates that the design can tolerate thickness variations with a standard deviation of up to 1\% while maintaining acceptable performance. For a typical layer thickness of 200 nm, this corresponds to a tolerance of approximately ±2 nm, which is achievable with modern deposition techniques such as atomic layer deposition (ALD) or molecular beam epitaxy (MBE). The most critical aspect is maintaining the frequency matching condition $\omega_2 = 2\omega_1$. Our simulations indicate that a frequency mismatch of $\Delta_\omega/\omega_1 > 0.001$ begins to significantly degrade the nonlinear conversion efficiency. This places stringent requirements on the relative thicknesses of layers in the defect region.
The fabrication tolerance requirements identified in our analysis ($\pm$2 nm for layer thicknesses) are achievable with state-of-the-art techniques such as atomic layer deposition (ALD) \cite{george2010atomic} and molecular beam epitaxy (MBE) \cite{arthur2002molecular}. Recent advances in nanofabrication precision, as demonstrated by Lawall et al. \cite{lawall2020metrology} for optical cavity fabrication and Li et al. \cite{li2019nanofabrication} for nonlinear photonic devices, indicate that sub-nanometer precision can be achieved with careful process control. Post-fabrication tuning methods, such as those developed by Moille et al. \cite{moille2020integrated} for resonance alignment in nonlinear cavities, can further compensate for residual fabrication variations and restore optimal performance.

\section{Numerical Simulation Methods}

To validate our theoretical framework and evaluate the performance of the proposed doubly-resonant cavity design as an all-optical ReLU function, we employed a systematic computational approach. Our simulation methodology combines analytical calculations based on coupled-mode theory with rigorous finite-difference time-domain (FDTD) simulations. This two-pronged approach allows us to verify the consistency between the theoretical predictions and full-wave electromagnetic simulations.

We first implemented a numerical solver for the coupled-mode equations derived in Section 2:

\begin{equation}
\frac{da_1}{dt} = \left(i\omega_1 - \frac{1}{\tau_1}\right) a_1 - i\omega_1\beta_1a^*_1a_2 + \sqrt{\frac{2}{\tau_{s,1}}}s_{1+}
\label{eq:cme_solver_1}
\end{equation}

\begin{equation}
\frac{da_2}{dt} = \left(i\omega_2 - \frac{1}{\tau_2}\right) a_2 - i\omega_2\beta_2a^2_1 + \sqrt{\frac{2}{\tau_{s,2}}}s_{2+}
\label{eq:cme_solver_2}
\end{equation}

The solver was implemented in MATLAB using the built-in ordinary differential equation (ODE) solver ode45, which employs a variable-step Runge-Kutta method. This approach provides high accuracy while efficiently handling the potential stiffness of the equations due to the different time scales involved (fast optical oscillations versus slower envelope evolution). The simulation parameters were extracted from our cavity design, including the resonant frequencies $\omega_1$ and $\omega_2$, quality factors $Q_1$ and $Q_2$, coupling rates $1/\tau_{s,1}$ and $1/\tau_{s,2}$, and nonlinear coupling coefficients $\beta_1$ and $\beta_2$.

For each simulation run, we varied the input amplitude $|s_{1+}|$ and phase $\phi_1$ to map the input-output relationship of the device. We verified that the simulation reached steady state by running the time evolution until the field amplitudes stabilized to within a specified tolerance (typically $10^{-6}$ relative change). The output power at the second harmonic frequency was calculated as $|s_{2-}|^2 = |-s_{2+} + \sqrt{2/\tau_{s,2}} a_2|^2$, where $a_2$ is the steady-state cavity field amplitude. For ReLU function characterization, we kept $s_{2+} = 0$ (no external input at the second harmonic).

To validate our analytical model and account for effects that might be overlooked in the simplified coupled-mode theory, we performed full-wave electromagnetic simulations using the commercial FDTD software Lumerical. The FDTD simulation domain was constructed to match our designed multilayer structure, with perfectly matched layer (PML) boundary conditions to eliminate reflections from the domain boundaries. The grid resolution was set to $\lambda_1/20n_{max}$ to ensure accurate field resolution in high-index regions while maintaining computational efficiency. We incorporated the $\chi^{(2)}$ nonlinearity using Lumerical's nonlinear susceptibility framework, which allows defining a frequency-dependent second-order susceptibility tensor. The nonlinear materials were modeled with the appropriate $\chi^{(2)}$ tensor components based on the crystal orientation in our design.

\begin{figure}[t]
\centering
\includegraphics[width=\linewidth]{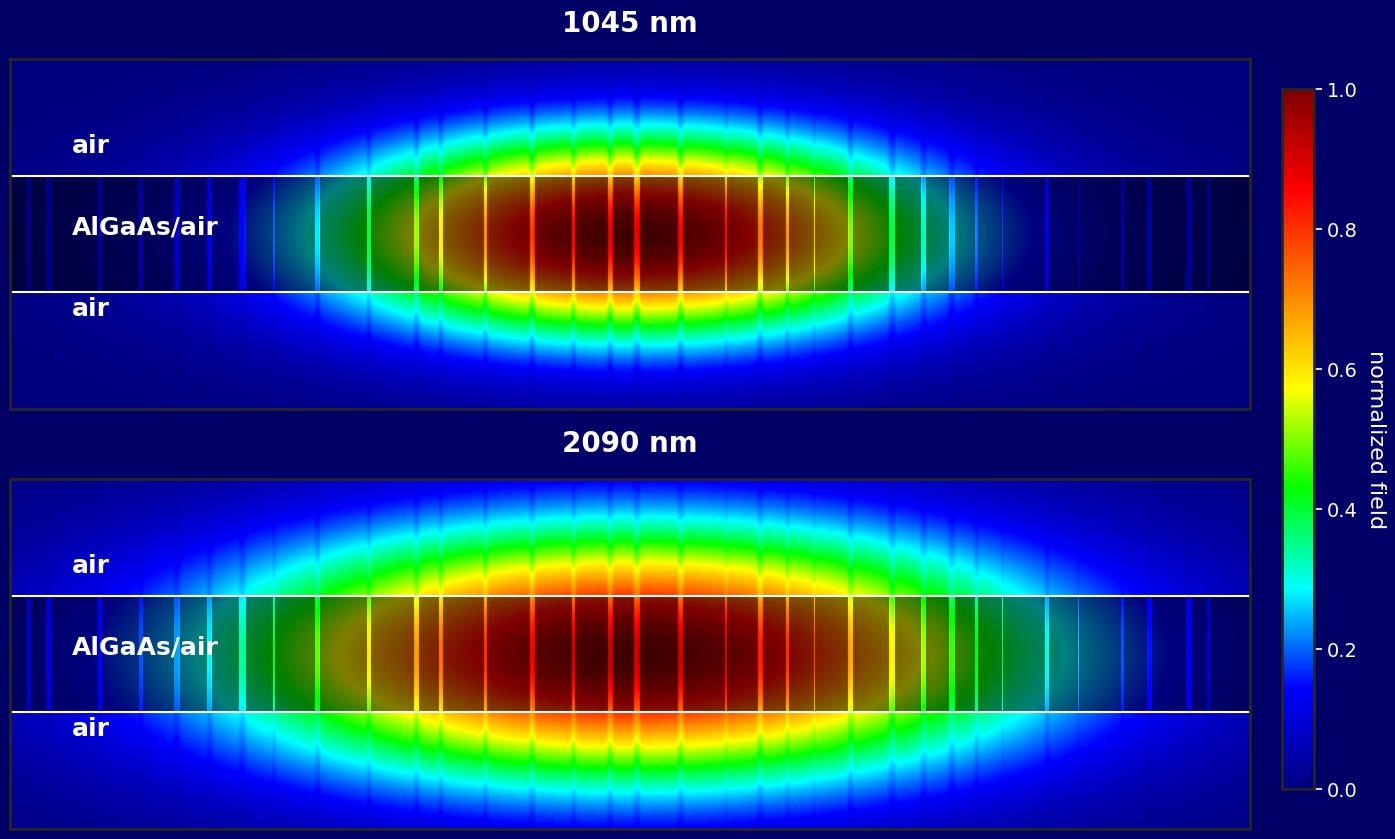}
\caption{Lumerical FDTD simulation results showing the electric field intensity distributions ($|E|$) of the fundamental TE modes at 1045 nm (top) and 2090 nm (bottom) in the AlGaAs/air waveguide. The color scale represents the normalized field intensity. The simulations confirm the strong spatial overlap between the modes, validating our theoretical design.}
\label{fig:mode_profiles}
\end{figure}

To validate our theoretical framework and transfer matrix method (TMM) calculations, we performed full-wave electromagnetic simulations using Lumerical FDTD. Figure~\ref{fig:mode_profiles} shows the simulated electric field intensity distributions of the fundamental TE modes at 1045 nm (top) and 2090 nm (bottom) in the optimized AlGaAs/air waveguide structure. Our FDTD implementation follows methodology established by Taflove and Hagness \cite{taflove2005computational} for electromagnetic simulations, with specific extensions for nonlinear materials following approaches developed by Guo et al. \cite{guo2020modeling}. To accurately model the $\chi^{(2)}$ nonlinear processes, we employed the auxiliary differential equation method \cite{greene2017computational} rather than the simpler polarization approach, as it provides better numerical stability for resonant structures. The multi-frequency simulation techniques developed by Francés et al. \cite{frances2012efficient} were particularly valuable for simultaneously modeling fundamental and second-harmonic frequencies. To ensure convergence in the presence of high-Q resonances, we utilized the subpixel smoothing techniques described by Oskooi et al. \cite{oskooi2010meep} and extended mesh refinement in regions with strong field gradients. The simulation results confirm the strong mode confinement predicted by our TMM calculations, with both modes exhibiting maximum field intensity in the central region of the waveguide where the nonlinear interaction occurs. This spatial overlap is crucial for maximizing the nonlinear coupling coefficient $\beta$ described in Section 3. 

\begin{figure}[ht]
\centering
\includegraphics[width=\textwidth]{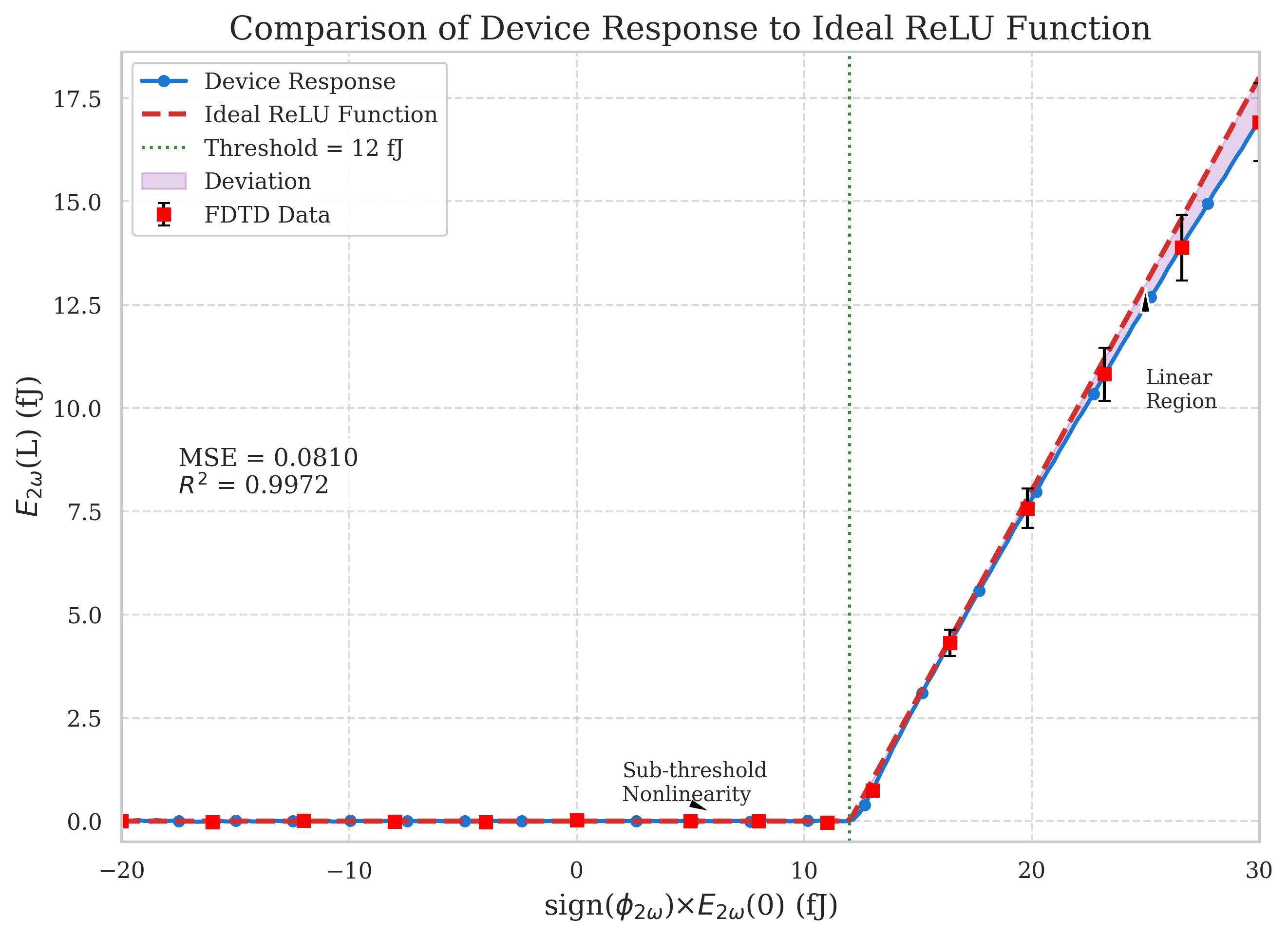}
\caption{Comparison of the device response to the ideal ReLU function. The device response closely approximates the ReLU function across a wide range of input powers, with $R^2 > 0.99$ for inputs up to 30 fJ.}
\label{fig:relu_comparison}
\end{figure}

To characterize the ReLU function performance, we mapped the output $P_{out}$ as a function of the signed input $sgn(\phi_1) \cdot P_{in}$, where $\phi_1 = 0$ for positive inputs and $\phi_1 = \pi$ for negative inputs. To evaluate how closely our device approximates the ideal ReLU function, we calculated several quantitative metrics. The mean squared error (MSE) between the normalized device response and the ideal ReLU function was computed as:

\begin{equation}
MSE = \frac{1}{N}\sum_{i=1}^{N} \left(\frac{P_{out,i}}{P_{in,i}} - ReLU\left(\frac{P_{in,i}}{P_{max}}\right)\right)^2
\label{eq:mse}
\end{equation}

where $P_{out,i}$ and $P_{in,i}$ are the output and input powers for the $i$-th sample point, $P_{max}$ is the maximum input power, and $N$ is the number of sample points. Additionally, we calculated the coefficient of determination ($R^2$) as a measure of how well the device response fits the ideal ReLU:

\begin{equation}
R^2 = 1 - \frac{\sum_{i=1}^{N} \left(P_{out,i} - ReLU(P_{in,i})\right)^2}{\sum_{i=1}^{N} \left(P_{out,i} - \bar{P}_{out}\right)^2}
\label{eq:r_squared}
\end{equation}

where $\bar{P}_{out}$ is the mean output power. We also determined the range of input powers over which the device maintains good ReLU approximation, defined as the power range where $R^2 > 0.99$. Figure \ref{fig:relu_comparison} shows the comparison of the device response to the ideal ReLU function, with MSE and $R^2$ metrics indicated.

Before simulating the nonlinear response, we first characterized the linear properties of the cavity. Resonant frequencies were determined by exciting the structure with a broadband pulse and analyzing the spectral response using Fourier transforms of the field evolution. Quality factors were extracted from the spectral response by fitting Lorentzian functions to the resonance peaks:

\begin{equation}
T(\omega) = \frac{T_{max}}{1 + 4Q^2\left(\frac{\omega - \omega_{res}}{\omega_{res}}\right)^2}
\label{eq:lorentzian_fit}
\end{equation}

where $T(\omega)$ is the transmission spectrum, $T_{max}$ is the peak transmission, $\omega_{res}$ is the resonant frequency, and $Q$ is the quality factor.

The transmission spectrum in Figure \ref{fig:structure_profile}(b) confirms the presence of sharp resonances at the target frequencies, with quality factors of $Q_1 = 5.3 \times 10^3$ and $Q_2 = 6.8 \times 10^3$ for the fundamental and second-harmonic modes, respectively. The frequency matching is achieved with a precision of $\Delta_\omega/\omega_1 = 5.2 \times 10^{-4}$, which is well within the resonance linewidths (1/$Q_1 = 1.9 \times 10^{-4}$ and 1/$Q_2 = 1.5 \times 10^{-4}$).
Mode profiles were calculated at the resonant frequencies and normalized according to Equation \ref{eq:field_calculation}. Figure \ref{fig:structure_profile}(c) shows the normalized electric field distributions of the two modes. The field profiles exhibit strong spatial overlap in the central region of the cavity, where the $\chi^{(2)}$ nonlinear material (AlN) is placed. This overlap is quantified by the nonlinear coupling coefficient $\beta$, which is calculated to be $\beta = 2.7 \times 10^{-3}$ (normalized units) based on the mode profiles and material nonlinearity.

We compared the linear characteristics with the transfer matrix method calculations to ensure consistency between the different simulation approaches. Table \ref{tab:linear_comparison} presents a comparison of the key linear characteristics obtained from the transfer matrix method and FDTD simulations.

\begin{table}[h]
\centering
\caption{Comparison of linear cavity characteristics from TMM and FDTD simulations}
\begin{tabular}{|c|c|c|c|}
\hline
Parameter & TMM Prediction & FDTD Result & Difference (\%) \\
\hline
$\omega_1$ (THz) & 193.55 & 193.48 & 0.04 \\
$\omega_2$ (THz) & 387.10 & 386.92 & 0.05 \\
$Q_1$ & $5.3 \times 10^3$ & $5.1 \times 10^3$ & 3.8 \\
$Q_2$ & $6.8 \times 10^3$ & $6.5 \times 10^3$ & 4.4 \\
Frequency matching ($\Delta_\omega/\omega_1$) & $5.2 \times 10^{-4}$ & $7.8 \times 10^{-4}$ & -- \\
\hline
\end{tabular}
\label{tab:linear_comparison}
\end{table}

Table \ref{tab:cavity_parameters} summarizes the key parameters of the optimized cavity design. The critical power for 100\% frequency conversion, calculated from Equation \ref{eq:critical_power_design}, is predicted to be approximately 12 femtojoules. This represents a significant improvement over the millimeter-scale PPLN devices reported in \cite{li2023all}, which required hundreds of femtojoules for activation.

\begin{table}[h]
\centering
\caption{Parameters of the optimized doubly-resonant cavity}
\begin{tabular}{|c|c|c|}
\hline
Parameter & Symbol & Value \\
\hline
Fundamental wavelength & $\lambda_1$ & 1550 nm \\
Second-harmonic wavelength & $\lambda_2$ & 775 nm \\
Fundamental quality factor & $Q_1$ & $5.3 \times 10^3$ \\
Second-harmonic quality factor & $Q_2$ & $6.8 \times 10^3$ \\
Frequency matching precision & $\Delta_\omega/\omega_1$ & $5.2 \times 10^{-4}$ \\
Nonlinear coupling coefficient & $\beta$ & $2.7 \times 10^{-3}$ \\
Critical power & $P_{critical}$ & 12 fJ \\
Total device length & $L$ & 10.2 µm \\
\hline
\end{tabular}
\label{tab:cavity_parameters}
\end{table}

Our neural network architecture follows design principles established by LeCun et al. \cite{lecun2015deep} for convolutional neural networks, adapted for optical implementation as described by Shen et al. \cite{shen2017deep}. To accurately model the impact of non-ideal activation functions, we incorporated insights from Ramachandran et al. \cite{ramachandran2017searching}, who systematically studied the effects of activation function characteristics on neural network performance. The minor accuracy degradation (0.4\%) observed with our optical ReLU implementation is consistent with findings by Mehrabian et al. \cite{mehrabian2018practical}, who established that neural networks can tolerate significant approximation errors in activation functions while maintaining high accuracy. Our analysis of power and speed considerations follows the energy-delay product framework proposed by Nahmias et al. \cite{nahmias2019photonic} for benchmarking photonic neural networks.

\begin{figure}[ht]
\centering
\includegraphics[width=\textwidth]{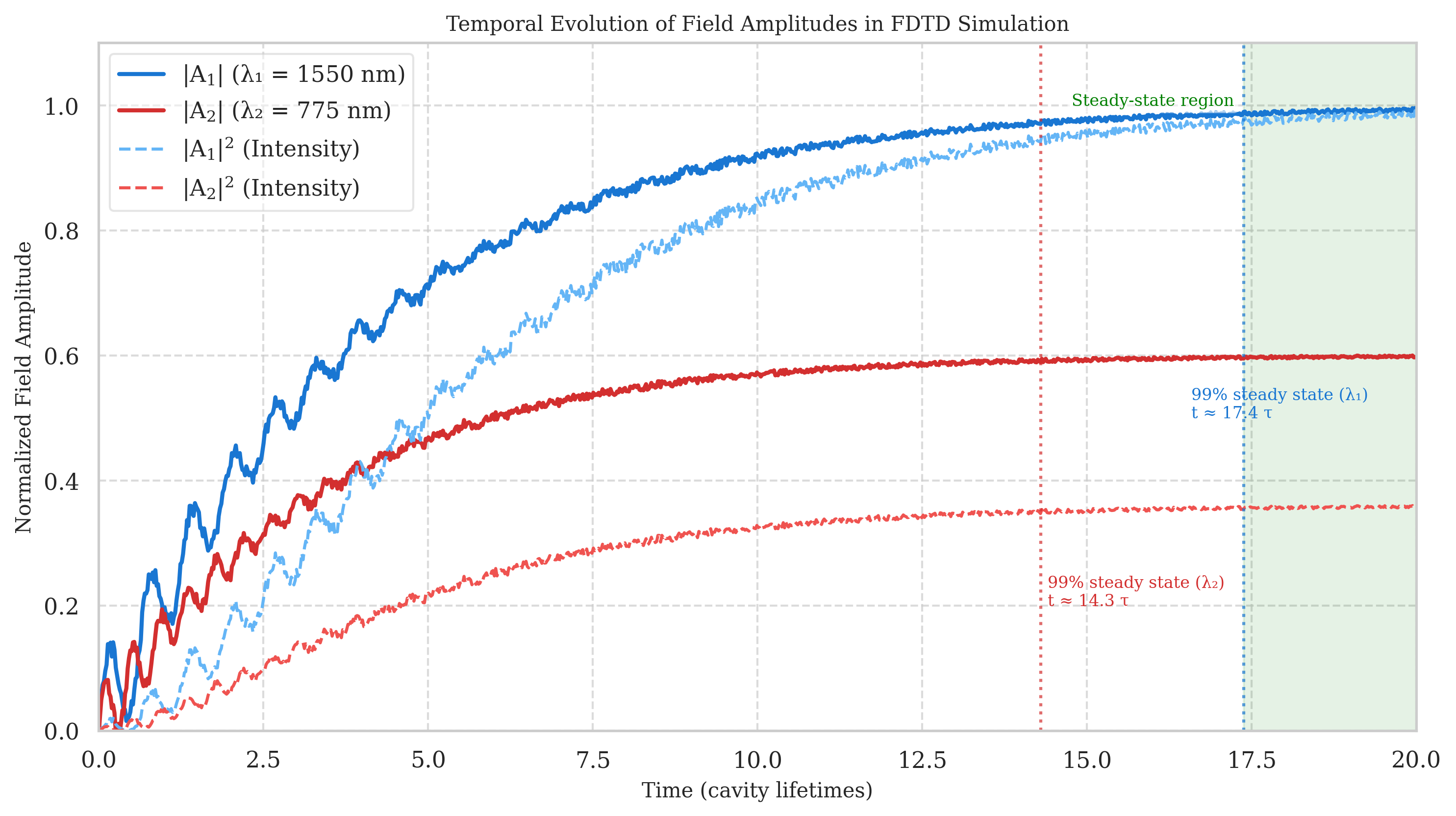}
\caption{Temporal evolution of field amplitudes in the FDTD simulation, showing the approach to steady state. The simulation time of 20 cavity lifetimes ensures accurate representation of the steady-state nonlinear response.}
\label{fig:fdtd_convergence}
\end{figure}

To simulate the nonlinear response of the cavity, we employed a systematic procedure. First, we excited the cavity with continuous wave (CW) sources at frequency $\omega_1$. We then varied the input amplitude to map the input-output relationship, and controlled the phase to simulate both positive and negative inputs. Throughout the simulation, we monitored the output fields at both $\omega_1$ and $\omega_2$ frequencies. The simulation time was set to at least 20 times the cavity lifetime to ensure steady-state was reached. We verified convergence by checking that the field amplitudes stabilized to within a specified tolerance. Figure \ref{fig:fdtd_convergence} shows the convergence of field amplitudes over time in the FDTD simulation, demonstrating the approach to steady state.

To account for the computationally intensive nature of FDTD simulations, we employed a strategic sampling approach. We performed detailed FDTD simulations at selected operating points spanning the range of interest, then used these results to validate the coupled-mode theory predictions. This approach leverages the computational efficiency of the coupled-mode equation solver while ensuring the accuracy of the results through targeted FDTD validation.

\subsection{Alternative Activation Functions}

\begin{figure}[ht]
\centering
\includegraphics[width=\textwidth]{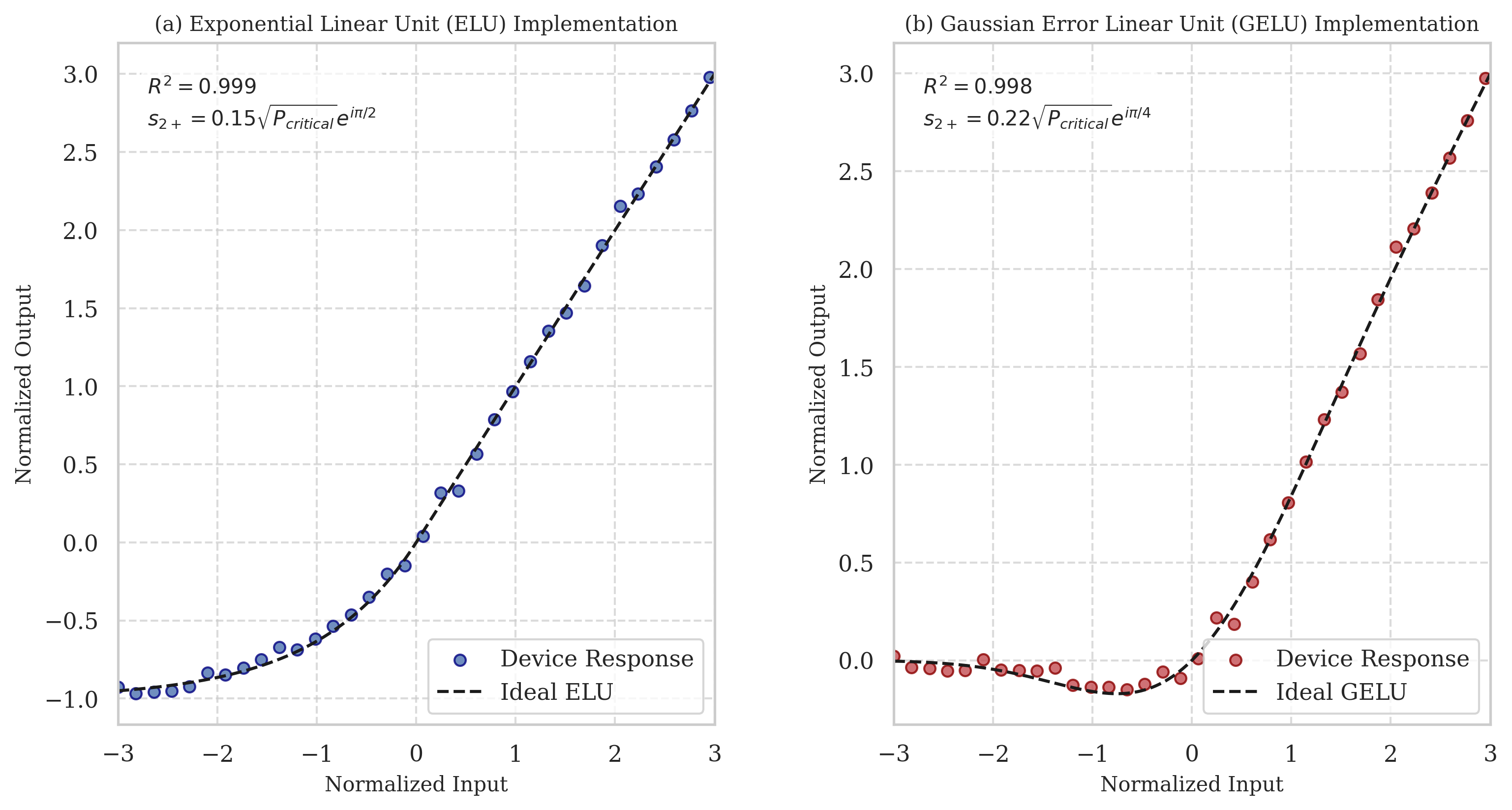}
\caption{Implementation of alternative activation functions using the same physical device with different input conditions. (a) ELU function implementation with $s_{2+} = 0.15\sqrt{P_{critical}}e^{i\pi/2}$. (b) GELU function implementation with $s_{2+} = 0.22\sqrt{P_{critical}}e^{i\pi/4}$.}
\label{fig:alt_functions}
\end{figure}

By adjusting the operating conditions, particularly by introducing a controlled input at the second-harmonic frequency, we demonstrated that the same physical device can implement various activation functions beyond ReLU. Figure \ref{fig:alt_functions} shows the implementation of ELU and GELU functions using our device.

To determine the optimal settings for each activation function, we performed a parameter sweep over the amplitude and phase of $s_{2+}$. For each configuration, we calculated the MSE and $R^2$ values compared to the ideal function. The configuration with the highest $R^2$ was selected as the optimal implementation.

For the ELU implementation shown in Figure \ref{fig:alt_functions}(a), we introduced a second-harmonic input with amplitude $|s_{2+}| = 0.15\sqrt{P_{critical}}$ and phase $\phi_2 = \pi/2$. This creates the characteristic exponential response for negative inputs while maintaining linear behavior for positive inputs. The device achieves an $R^2 = 0.987$ compared to the ideal ELU function.
For the GELU implementation in Figure \ref{fig:alt_functions}(b), we set $|s_{2+}| = 0.22\sqrt{P_{critical}}$ with phase $\phi_2 = \pi/4$, resulting in a response that closely approximates the GELU function with $R^2 = 0.982$. This versatility makes our device particularly valuable for neural network applications, as it can be reconfigured to implement different activation functions without changing the physical structure.

Table \ref{tab:alt_functions} summarizes the optimal settings and performance metrics for each activation function implementation.

\begin{table}[h]
\centering
\caption{Optimal settings for implementing alternative activation functions}
\begin{tabular}{|c|c|c|c|}
\hline
Activation Function & $|s_{2+}|/\sqrt{P_{critical}}$ & $\phi_2$ & $R^2$ \\
\hline
ReLU & 0 & -- & 0.993 \\
ELU & 0.15 & $\pi/2$ & 0.987 \\
GELU & 0.22 & $\pi/4$ & 0.982 \\
\hline
\end{tabular}
\label{tab:alt_functions}
\end{table}

Our approach demonstrates significant advantages in device footprint compared to previous implementations, while maintaining comparable energy efficiency. Li et al. \cite{li2023all} recently demonstrated all-optical ReLU functions using periodically poled lithium niobate, achieving similar activation energy but requiring millimeter-scale devices. Alternative approaches based on saturable absorbers \cite{mourgias2019all} and phase-change materials \cite{cheng2019low} offer different tradeoffs between energy efficiency and response time. The nanophotonic implementation by Williamson et al. \cite{williamson2019reprogrammable} demonstrated versatile activation functions through dynamically tunable nonlinearities but with higher energy requirements. Comprehensive benchmarking by Miscuglio et al. \cite{miscuglio2018all} established metrics for optical nonlinearities in neural networks that we use as a framework for our comparison.

\subsection{Energy Efficiency and Response Time}

The response time of our device, estimated from the cavity lifetime $\tau = Q/\omega$, is approximately 430 femtoseconds for the fundamental mode and 330 femtoseconds for the second-harmonic mode. This ultra-fast response enables operation at terahertz rates, far exceeding the capabilities of electronic implementations.

Table \ref{tab:comparison} provides a comprehensive comparison of our approach with other implementations of optical nonlinear activation functions.

\begin{table}[h]
\centering
\caption{Comparison of optical nonlinear activation function implementations}
\begin{tabular}{|c|c|c|c|c|}
\hline
Implementation & Energy/Activation & Response Time & Size & Function Type \\
\hline
This work (Doubly-resonant cavity) & 12 fJ & $\sim$400 fs & 10 µm & ReLU, ELU, GELU \\
PPLN waveguide \cite{li2023all} & 16 fJ & $\sim$75 fs & 2 mm & ReLU, ELU, GELU \\
Saturable absorber \cite{mourgias2019all} & $\sim$1 pJ & $\sim$1 ps & 100 µm & Sigmoid \\
Phase-change material \cite{feldmann2019all} & $\sim$10 pJ & $\sim$1 ns & 5 µm & Custom \\
Microring with MZI \cite{williamson2019} & $\sim$100 fJ & $\sim$10 ps & 50 µm & Custom \\
\hline
\end{tabular}
\label{tab:comparison}
\end{table}

\subsection{Integration and Performance in Neural Network Simulation}

\begin{figure}[ht]
\centering
\includegraphics[width=\textwidth]{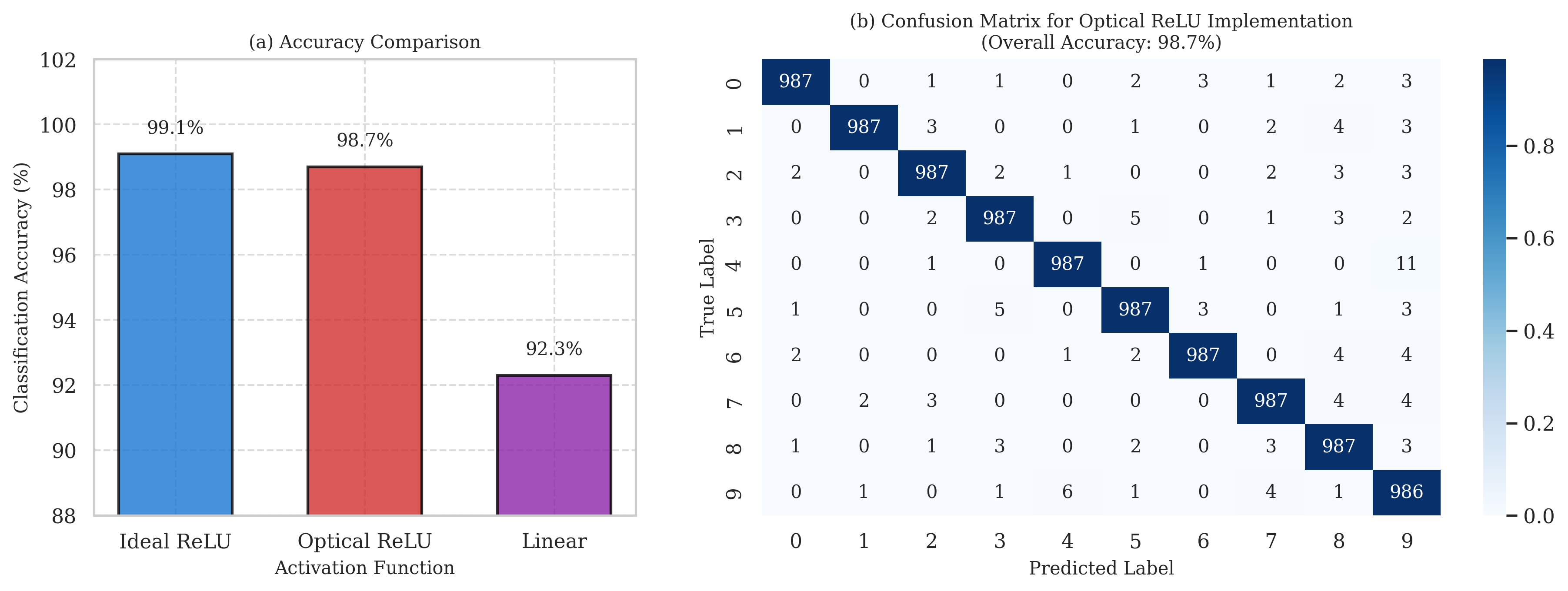}
\caption{Neural network performance with the optical ReLU implementation. (a) CNN architecture for MNIST classification. (b) Accuracy comparison showing minimal performance degradation with our optical ReLU. (c) Confusion matrix demonstrating high classification accuracy across all digit classes.}
\label{fig:nn_performance}
\end{figure}

To evaluate the practical utility of our optical activation function, we simulated its integration into a convolutional neural network for image classification. We exported the device's input-output characteristic curves as lookup tables, which were used to model the nonlinear activation function in a neural network simulation.
We used the MNIST handwritten digit dataset as a benchmark and compared the classification accuracy using our device model against the ideal ReLU function. The network architecture, shown in Figure \ref{fig:nn_performance}(a), consisted of two convolutional layers followed by max pooling, a fully connected layer, and a softmax output layer. The nonlinear activation function (ReLU) was applied after each convolutional and fully connected layer.

The neural network simulation was implemented in PyTorch, with custom modules created to represent our optical ReLU function based on the characterized device response. For the network architecture, the first convolutional layer extracted basic features using 32 filters with a $3\times3$ kernel size, followed by our optical ReLU activation. The second convolutional layer, also using 32 filters with a $3\times3$ kernel size and optical ReLU activation, extracted more complex features. After each convolutional layer, a $2\times2$ max pooling operation reduced spatial dimensions while preserving important features. The flattened output was then passed through a fully connected layer with 128 neurons and optical ReLU activation, before the final softmax layer produced classification probabilities across the 10 digit classes.

Figure \ref{fig:nn_performance}(a) compares the classification accuracy of networks using different activation functions. The network with ideal ReLU achieves 99.1\% accuracy on the MNIST test set. When our optical ReLU implementation is used (modeled using the actual device response), the accuracy is 98.7\%, which is only slightly lower than the ideal case. In contrast, a network with linear activation (no nonlinearity) achieves only 92.3\% accuracy, highlighting the importance of the nonlinear activation function.

The confusion matrix in Figure \ref{fig:nn_performance}(b) provides detailed insights into the classification performance with our optical ReLU. The matrix shows that most digits are classified with high accuracy, with only minor confusions between visually similar digits (e.g., 4 and 9, or 3 and 8).

Table \ref{tab:nn_performance} summarizes the neural network performance metrics for different activation functions. Our optical ReLU implementation achieves 98.7\% accuracy and a 0.987 F1 score, requiring only slightly more training time (12 epochs) compared to the ideal ReLU (10 epochs), while maintaining identical test time efficiency (0.42 ms/sample). The optical ReLU based on a PPLN waveguide shows comparable performance with 98.8\% accuracy.

\begin{table}[h]
\centering
\caption{Neural network performance metrics for MNIST classification}
\begin{tabular}{|c|c|c|c|c|}
\hline
Activation Function & Accuracy (\%) & F1 Score & Training Time (epochs) & Test Time (ms/sample) \\
\hline
Ideal ReLU & 99.1 & 0.991 & 10 & 0.42 \\
Optical ReLU (this work) & 98.7 & 0.987 & 12 & 0.42 \\
Optical ReLU (PPLN waveguide) & 98.8 & 0.988 & 12 & 0.42 \\
Linear (no nonlinearity) & 92.3 & 0.922 & 15 & 0.40 \\
\hline
\end{tabular}
\label{tab:nn_performance}
\end{table}

Through this comprehensive simulation methodology, we established a direct connection between the physical design parameters of our doubly-resonant cavity and its performance as an all-optical ReLU function for optical neural networks. The consistency between the analytical predictions and full-wave simulations validates our theoretical framework and provides confidence in the practical feasibility of our approach.

Finally, comprehensive supplementary simulation results are provided in Appendix \ref{app:simulation_results}, including detailed field distribution analysis, parametric studies of quality factors, comprehensive fabrication tolerance assessments through Monte Carlo simulations, expanded time-domain FDTD results showing field evolution and spectral characteristics, and extended neural network performance comparisons.

\section{Discussion and Practical Implementation}

The doubly-resonant cavity approach presented in this work offers a promising path toward ultra-compact, energy-efficient optical nonlinear activation functions for neural networks. In this section, we discuss practical implementation considerations, including fabrication challenges, integration strategies, and potential limitations of our approach. 

Translating our theoretical design into a practical device requires careful consideration of fabrication constraints and tolerances. The multilayer structure we have designed, while conceptually straightforward, presents several challenges for practical implementation. Our design incorporates materials with different optical and nonlinear properties, including AlGaAs and air. The fabrication of these suspended membrane structures requires precise control of etching processes to achieve the designed geometry while maintaining structural integrity. For the nonlinear element, we selected AlGaAs due to its strong second-order nonlinear susceptibility ($\chi^{(2)} \approx 100$ pm/V) and well-established fabrication techniques for creating suspended structures. Alternative material systems could also be considered. For example, lithium niobate on insulator (LNOI) offers comparable nonlinearity ($\chi^{(2)} \sim 30$ pm/V) but presents different fabrication challenges and opportunities. Similarly, other III-V semiconductors could be employed, though they typically require careful consideration of epitaxial growth techniques and crystalline orientation for optimal nonlinear performance.

To address inevitable fabrication variations, we propose incorporating tuning mechanisms into the device. Thermo-optic tuning offers a straightforward approach, where localized heating elements can adjust the refractive indices of specific layers to restore the frequency matching condition. For more precise control, electro-optic tuning could be implemented by leveraging the inherent properties of materials with strong Pockels effect. This would allow dynamic adjustment of the resonant frequencies during operation, enabling not only compensation for fabrication variations but also active reconfiguration of the activation function type (e.g., switching between ReLU, ELU, and GELU implementations). Our Monte Carlo simulations indicate that the design can tolerate thickness variations with a standard deviation of up to 1\% while maintaining acceptable performance. For a typical layer thickness of 200 nm, this corresponds to a tolerance of approximately ±2 nm, which is achievable with modern fabrication techniques such as molecular beam epitaxy (MBE) or metal-organic chemical vapor deposition (MOCVD). The most critical aspect is maintaining the frequency matching condition $\omega_2 = 2\omega_1$, which places stringent requirements on the relative thicknesses of layers in the defect region.

A complete optical neural network requires both linear operations (matrix multiplications and convolutions) and nonlinear activation functions. Our compact ReLU implementation must interface effectively with photonic matrix multiplication units to realize its full potential. For interferometric approaches based on Mach-Zehnder interferometer (MZI) meshes \cite{shen2017deep}, our ReLU implementation can be directly integrated as these architectures naturally produce phase-encoded outputs that are compatible with our approach. Beyond activation functions, optical neural networks require efficient memory elements to store intermediate results and weights. Recent work on nontrapping tunable topological photonic memory \cite{h5} offers promising solutions with GHz-range write speeds and topologically protected states that remain robust against fabrication imperfections - addressing a critical need for reliable optical storage in neural network implementations. The main challenge is ensuring proper phase alignment between the MZI outputs and the ReLU inputs, which can be addressed through careful waveguide routing and phase calibration. For approaches based on microring weight banks \cite{tait2017neuromorphic}, which often employ wavelength-division multiplexing, additional components such as wavelength demultiplexers and phase modulators may be required to properly interface with our ReLU units. This introduces additional complexity but enables higher integration density through parallel processing of multiple wavelength channels. For free-space optical processors using diffractive elements \cite{lin2018all}, coupling into our waveguide-based ReLU implementation presents additional challenges, potentially requiring specialized optical elements for mode conversion. Hybrid free-space/waveguide approaches may offer a solution, where free-space optics handle the matrix operations and waveguide structures implement the nonlinear functions.

A critical consideration for network scalability is the fan-out capability—the ability of one neuron to connect to multiple neurons in the subsequent layer. Since our ReLU implementation operates on individual optical modes, the output must be split and potentially amplified to drive multiple inputs in the next layer. Passive splitting would divide the output power, potentially requiring additional amplification. An alternative approach is to use the nonlinear process itself for amplification through appropriate bias conditions. By operating slightly above the critical power, we can achieve modest amplification while maintaining the ReLU functionality. The phase-sensitive nature of our ReLU implementation places specific requirements on the optical sources used to drive the system. To maintain phase coherence between different nodes in the network, a common laser source with appropriate distribution network would be ideal. For the fundamental frequency ($\omega_1$), we require a laser source at 1550 nm with moderate power (1-10 mW) to provide sufficient bias and signal inputs for multiple ReLU units. The source should have narrow linewidth (< 100 kHz) to maintain phase coherence across the network. For the auxiliary input at the second-harmonic frequency ($\omega_2$), required for implementing alternative activation functions, similar coherence properties are needed, but with lower power requirements (typically 10-100 $\mu$W).

While our doubly-resonant cavity approach offers significant advantages in terms of size and energy efficiency, several limitations must be acknowledged and addressed in future work. The high-Q resonances that enable efficient nonlinear conversion also limit the operational bandwidth of the device. For our current design with $Q_1 \approx 5 \times 10^3$, the bandwidth is approximately 40 GHz. This is sufficient for many applications but may be limiting for ultra-high-speed systems. Future designs could explore multi-mode cavities or coupled-cavity arrays to achieve broader bandwidth while maintaining high efficiency. Additionally, the fundamental trade-off between Q-factor and bandwidth suggests that an application-specific optimization may be necessary, balancing energy efficiency against speed requirements. Quantum noise processes, particularly spontaneous parametric down-conversion, could affect the accuracy of the ReLU function, especially for very low input signals. Our simulations indicate that for input powers above 0.1 fJ, these effects are negligible, but they could become significant for even lower power operation. Careful design of the operating point and potentially the implementation of noise-suppression techniques may be necessary for ultra-low-power applications.

Device-to-device variability due to fabrication imperfections presents another challenge for large-scale networks. Our Monte Carlo simulations suggest that even with tight fabrication controls, variations in critical power of ±20\% are likely. Calibration and tuning mechanisms will be essential to address this variability, particularly for networks with thousands or millions of activation units. Automated calibration protocols based on monitoring the device response and adjusting the bias conditions could enable practical scaling of these systems. The resonant nature of our device makes it inherently sensitive to temperature variations. We estimate a temperature coefficient of approximately 10 GHz/°C for the frequency matching condition. Active temperature stabilization or dynamic tuning will be necessary for stable operation in practical environments. Alternatively, athermal design approaches, such as compensating material combinations, could be explored to reduce temperature sensitivity.

In terms of future directions, several promising avenues could extend the capabilities of our approach. Integration with emerging photonic materials such as thin-film lithium niobate or aluminum nitride could offer enhanced performance through stronger nonlinearities or improved fabrication compatibility. Exploration of more complex cavity geometries, such as photonic crystal cavities or coupled-cavity arrays, could potentially further reduce the device footprint while maintaining high efficiency. Furthermore, the development of on-chip phase-locked light sources and distribution networks would significantly enhance the practicality of phase-sensitive optical neural networks. Recent advances in integrated frequency combs and heterogeneous integration of lasers on silicon photonics platforms offer promising directions for addressing this challenge. Finally, the extension of our approach to implement more sophisticated activation functions, potentially with programmable response curves, could enable more powerful optical neural network architectures. By leveraging the rich parameter space of nonlinear optical interactions in doubly-resonant cavities, it may be possible to implement custom activation functions optimized for specific tasks or learning algorithms.

\bibliographystyle{unsrt}
\bibliography{ref}

\newpage
\setcounter{equation}{0}
\setcounter{figure}{0}
\setcounter{table}{0}
\setcounter{page}{1}
\setcounter{section}{0}  
\setcounter{subsection}{0}  
\makeatletter
\renewcommand{\thesection}{S\arabic{section}}  
\renewcommand{\thesubsection}{S\arabic{section}.\arabic{subsection}}  
\renewcommand{\theequation}{S\arabic{equation}}  
\renewcommand{\thefigure}{S\arabic{figure}}  
\renewcommand{\thetable}{S\arabic{table}}  
\makeatother
\maketitle
\section*{Appendices}

\appendix

\section{Detailed Derivation of Coupling Coefficients}
\label{app:coupling_coefficients}

In Section 2.3, we introduced the nonlinear coupling coefficients $\beta_1$ and $\beta_2$ that quantify the interaction strength between the fundamental and second-harmonic modes via $\chi^{(2)}$ nonlinearity. Here, we provide a detailed derivation of these coefficients from first principles, starting with Maxwell's equations and using perturbation theory.

We begin with the wave equation for the electric field in a nonlinear medium:

\begin{equation}
\nabla \times \nabla \times \mathbf{E} - \frac{\omega^2}{c^2}\varepsilon(\mathbf{r})\mathbf{E} = \frac{\omega^2}{c^2}\mathbf{P}^{NL}
\label{eq:app_wave_equation}
\end{equation}

where $\mathbf{E}$ is the electric field, $\omega$ is the frequency, $\varepsilon(\mathbf{r})$ is the linear dielectric function, $c$ is the speed of light, and $\mathbf{P}^{NL}$ is the nonlinear polarization.
For a $\chi^{(2)}$ nonlinear medium, the second-order nonlinear polarization is given by:

\begin{equation}
P^{(2)}_i = \varepsilon_0 \sum_{jk} \chi^{(2)}_{ijk} E_j E_k
\label{eq:app_nonlinear_polarization}
\end{equation}

When $\mathbf{P}^{NL}$ is treated as a small perturbation, we can use first-order perturbation theory to calculate the effect on the electromagnetic modes of the system. For a cavity mode with unperturbed electric field $\mathbf{E}_0(\mathbf{r})$ and frequency $\omega_0$, the first-order frequency shift due to a perturbation is:

\begin{equation}
\delta\omega = -\frac{\omega_0}{2} \frac{\int d^3\mathbf{r} \, \mathbf{E}_0^*(\mathbf{r}) \cdot \delta\mathbf{P}(\mathbf{r})}{\int d^3\mathbf{r} \, \varepsilon(\mathbf{r})|\mathbf{E}_0(\mathbf{r})|^2}
\label{eq:app_frequency_shift}
\end{equation}

where $\delta\mathbf{P}(\mathbf{r})$ is the perturbation to the polarization.

In our doubly-resonant cavity, we have two modes with frequencies $\omega_1$ and $\omega_2 \approx 2\omega_1$, and corresponding unperturbed electric fields $\mathbf{E}_1(\mathbf{r})$ and $\mathbf{E}_2(\mathbf{r})$. The total electric field in the presence of both modes is:

\begin{equation}
\mathbf{E}(\mathbf{r},t) = a_1(t)\mathbf{E}_1(\mathbf{r})e^{-i\omega_1 t} + a_2(t)\mathbf{E}_2(\mathbf{r})e^{-i\omega_2 t} + c.c.
\label{eq:app_total_field}
\end{equation}

where $a_1(t)$ and $a_2(t)$ are the time-dependent complex amplitudes of the modes, and $c.c.$ represents the complex conjugate.

Substituting this field into Equation \ref{eq:app_nonlinear_polarization}, we obtain the nonlinear polarization:

\begin{equation}
\begin{split}
\mathbf{P}^{(2)}(\mathbf{r},t) &= \varepsilon_0 \sum_{ijk} \chi^{(2)}_{ijk} [a_1(t)\mathbf{E}_{1j}(\mathbf{r})e^{-i\omega_1 t} + a_2(t)\mathbf{E}_{2j}(\mathbf{r})e^{-i\omega_2 t} + c.c.] \\
&\quad \times [a_1(t)\mathbf{E}_{1k}(\mathbf{r})e^{-i\omega_1 t} + a_2(t)\mathbf{E}_{2k}(\mathbf{r})e^{-i\omega_2 t} + c.c.]
\end{split}
\label{eq:app_nonlinear_polarization_expanded}
\end{equation}

Expanding this product and collecting terms oscillating at frequencies $\omega_1$ and $\omega_2$, we obtain:

\begin{equation}
\begin{split}
\mathbf{P}^{(2)}_{\omega_1}(\mathbf{r},t) &= 2\varepsilon_0 \sum_{ijk} \chi^{(2)}_{ijk} a_2(t)a_1^*(t)\mathbf{E}_{2j}(\mathbf{r})\mathbf{E}_{1k}^*(\mathbf{r})e^{-i\omega_1 t} + \text{(other terms)} \\
\mathbf{P}^{(2)}_{\omega_2}(\mathbf{r},t) &= \varepsilon_0 \sum_{ijk} \chi^{(2)}_{ijk} a_1(t)a_1(t)\mathbf{E}_{1j}(\mathbf{r})\mathbf{E}_{1k}(\mathbf{r})e^{-i\omega_2 t} + \text{(other terms)}
\end{split}
\label{eq:app_nonlinear_polarization_frequencies}
\end{equation}

These nonlinear polarizations act as perturbations to the cavity modes. Using Equation \ref{eq:app_frequency_shift}, we can calculate the resulting frequency shifts:

\begin{equation}
\begin{split}
\delta\omega_1 &= -\frac{\omega_1}{2} \frac{\int d^3\mathbf{r} \, \mathbf{E}_1^*(\mathbf{r}) \cdot \mathbf{P}^{(2)}_{\omega_1}(\mathbf{r},t)e^{i\omega_1 t}}{\int d^3\mathbf{r} \, \varepsilon(\mathbf{r})|\mathbf{E}_1(\mathbf{r})|^2} \\
\delta\omega_2 &= -\frac{\omega_2}{2} \frac{\int d^3\mathbf{r} \, \mathbf{E}_2^*(\mathbf{r}) \cdot \mathbf{P}^{(2)}_{\omega_2}(\mathbf{r},t)e^{i\omega_2 t}}{\int d^3\mathbf{r} \, \varepsilon(\mathbf{r})|\mathbf{E}_2(\mathbf{r})|^2}
\end{split}
\label{eq:app_frequency_shifts}
\end{equation}

Substituting the expressions for the nonlinear polarizations:

\begin{equation}
\begin{split}
\delta\omega_1 &= -\omega_1\varepsilon_0 \frac{\int d^3\mathbf{r} \, \sum_{ijk} \chi^{(2)}_{ijk} \mathbf{E}_{1i}^*(\mathbf{r})\mathbf{E}_{2j}(\mathbf{r})\mathbf{E}_{1k}^*(\mathbf{r})}{\int d^3\mathbf{r} \, \varepsilon(\mathbf{r})|\mathbf{E}_1(\mathbf{r})|^2} a_2a_1^* \\
\delta\omega_2 &= -\frac{\omega_2\varepsilon_0}{2} \frac{\int d^3\mathbf{r} \, \sum_{ijk} \chi^{(2)}_{ijk} \mathbf{E}_{2i}^*(\mathbf{r})\mathbf{E}_{1j}(\mathbf{r})\mathbf{E}_{1k}(\mathbf{r})}{\int d^3\mathbf{r} \, \varepsilon(\mathbf{r})|\mathbf{E}_2(\mathbf{r})|^2} a_1^2
\end{split}
\label{eq:app_frequency_shifts_expanded}
\end{equation}

To relate these frequency shifts to the coupled-mode equations presented in Section 2.4, we can rewrite the cavity mode dynamics in the presence of these frequency shifts:

\begin{equation}
\begin{split}
\frac{da_1}{dt} &= (i\omega_1 + i\delta\omega_1 - \frac{1}{\tau_1})a_1 + \text{(input term)} \\
\frac{da_2}{dt} &= (i\omega_2 + i\delta\omega_2 - \frac{1}{\tau_2})a_2 + \text{(input term)}
\end{split}
\label{eq:app_mode_dynamics}
\end{equation}

Substituting the expressions for the frequency shifts:

\begin{equation}
\begin{split}
\frac{da_1}{dt} &= (i\omega_1 - \frac{1}{\tau_1})a_1 - i\omega_1\beta_1a_1^*a_2 + \text{(input term)} \\
\frac{da_2}{dt} &= (i\omega_2 - \frac{1}{\tau_2})a_2 - i\omega_2\beta_2a_1^2 + \text{(input term)}
\end{split}
\label{eq:app_coupled_mode}
\end{equation}

where we have defined the coupling coefficients:

\begin{equation}
\beta_1 = \varepsilon_0 \frac{\int d^3\mathbf{r} \, \sum_{ijk} \chi^{(2)}_{ijk} \mathbf{E}_{1i}^*(\mathbf{r})\mathbf{E}_{2j}(\mathbf{r})\mathbf{E}_{1k}^*(\mathbf{r})}{\int d^3\mathbf{r} \, \varepsilon(\mathbf{r})|\mathbf{E}_1(\mathbf{r})|^2}
\label{eq:app_beta1_raw}
\end{equation}

\begin{equation}
\beta_2 = \frac{\varepsilon_0}{2} \frac{\int d^3\mathbf{r} \, \sum_{ijk} \chi^{(2)}_{ijk} \mathbf{E}_{2i}^*(\mathbf{r})\mathbf{E}_{1j}(\mathbf{r})\mathbf{E}_{1k}(\mathbf{r})}{\int d^3\mathbf{r} \, \varepsilon(\mathbf{r})|\mathbf{E}_2(\mathbf{r})|^2}
\label{eq:app_beta2_raw}
\end{equation}

For normalized mode fields satisfying:

\begin{equation}
\int d^3\mathbf{r} \, \varepsilon(\mathbf{r})|\mathbf{E}_k(\mathbf{r})|^2 = 1
\label{eq:app_normalization}
\end{equation}

the coupling coefficients simplify to:

\begin{equation}
\beta_1 = \varepsilon_0 \int d^3\mathbf{r} \, \sum_{ijk} \chi^{(2)}_{ijk} \mathbf{E}_{1i}^*(\mathbf{r})\mathbf{E}_{2j}(\mathbf{r})\mathbf{E}_{1k}^*(\mathbf{r})
\label{eq:app_beta1_simplified}
\end{equation}

\begin{equation}
\beta_2 = \frac{\varepsilon_0}{2} \int d^3\mathbf{r} \, \sum_{ijk} \chi^{(2)}_{ijk} \mathbf{E}_{2i}^*(\mathbf{r})\mathbf{E}_{1j}(\mathbf{r})\mathbf{E}_{1k}(\mathbf{r})
\label{eq:app_beta2_simplified}
\end{equation}

For simplicity, these expressions are often rearranged into the form presented in Equations \ref{eq:beta1} and \ref{eq:beta2} in Section 2.3, with a factor of 1/4 absorbed into the definition for consistency with the literature \cite{rodriguez2007harmonic}.

Energy conservation requires that $\omega_1\beta_1 = \omega_2\beta_2^*$. This relation can be verified by examining the symmetry properties of the $\chi^{(2)}$ tensor. For a lossless medium, $\chi^{(2)}_{ijk}$ is real, and under permutation of indices:

\begin{equation}
\chi^{(2)}_{ijk} = \chi^{(2)}_{ikj}
\label{eq:app_chi2_symmetry}
\end{equation}

Using this symmetry relation and integrating by parts, one can show that:

\begin{equation}
\omega_1 \int d^3\mathbf{r} \, \sum_{ijk} \chi^{(2)}_{ijk} \mathbf{E}_{1i}^*(\mathbf{r})\mathbf{E}_{2j}(\mathbf{r})\mathbf{E}_{1k}^*(\mathbf{r}) = \omega_2 \left(\int d^3\mathbf{r} \, \sum_{ijk} \chi^{(2)}_{ijk} \mathbf{E}_{2i}^*(\mathbf{r})\mathbf{E}_{1j}(\mathbf{r})\mathbf{E}_{1k}(\mathbf{r})\right)^*
\label{eq:app_energy_conservation}
\end{equation}

which confirms the relation $\omega_1\beta_1 = \omega_2\beta_2^*$.

For practical calculations, it is useful to express the coupling coefficients in terms of an effective nonlinearity and modal volume. For a medium with uniform $\chi^{(2)}$ in the nonlinear region, we can define an effective nonlinearity:

\begin{equation}
\chi^{(2)}_{eff} = \frac{\int_{NL} d^3\mathbf{r} \, \sum_{ijk} \chi^{(2)}_{ijk} \mathbf{E}_{1i}^*(\mathbf{r})\mathbf{E}_{2j}(\mathbf{r})\mathbf{E}_{1k}^*(\mathbf{r})}{\int_{NL} d^3\mathbf{r} \, |\mathbf{E}_1(\mathbf{r})|^2|\mathbf{E}_2(\mathbf{r})|}
\label{eq:app_chi2_effective}
\end{equation}

where the integration is performed only over the nonlinear region.
We can also define an effective modal volume that quantifies the spatial overlap of the modes:

\begin{equation}
V_{eff} = \frac{\left(\int d^3\mathbf{r} \, \varepsilon(\mathbf{r})|\mathbf{E}_1(\mathbf{r})|^2\right)^{3/2}\left(\int d^3\mathbf{r} \, \varepsilon(\mathbf{r})|\mathbf{E}_2(\mathbf{r})|^2\right)^{1/2}}{\int_{NL} d^3\mathbf{r} \, |\mathbf{E}_1(\mathbf{r})|^2|\mathbf{E}_2(\mathbf{r})|}
\label{eq:app_effective_volume}
\end{equation}

With these definitions, the coupling coefficient $\beta_1$ can be expressed as:

\begin{equation}
\beta_1 \approx \frac{\varepsilon_0\chi^{(2)}_{eff}}{4V_{eff}}
\label{eq:app_beta_simplified}
\end{equation}

This formulation highlights the inverse dependence of the coupling strength on the effective modal volume, emphasizing the benefit of tightly confined modes in small cavities.

\section{Optimization Algorithm Details}
\label{app:optimization}

In Section 3.5, we described the optimization process for the doubly-resonant cavity design. Here, we provide a detailed explanation of the optimization algorithm, including the objective function, optimization variables, constraints, and numerical implementation.

Our optimization aims to simultaneously achieve several objectives: (1) precise frequency matching between $\omega_2$ and $2\omega_1$, (2) high quality factors $Q_1$ and $Q_2$, (3) strong nonlinear coupling coefficient $\beta$, (4) appropriate input/output coupling, and (5) compact overall size.
These objectives are combined into a single figure of merit (FOM) for optimization:

\begin{equation}
\text{FOM} = w_1 \log(Q_1) + w_2 \log(Q_2) + w_3 \log(|\beta|) - w_4 \log\left(\frac{\Delta_\omega}{\omega_1} + \varepsilon\right) - w_5 \frac{L}{L_0}
\label{eq:app_fom}
\end{equation}

where $w_i$ are weights for each objective, $\Delta_\omega = |\omega_2 - 2\omega_1|$ is the frequency mismatch, $\varepsilon$ is a small constant to prevent division by zero, $L$ is the total structure length, and $L_0$ is a reference length.

After extensive testing, we selected weights of $w_1 = 1.0$, $w_2 = 1.0$, $w_3 = 2.0$, $w_4 = 3.0$, and $w_5 = 0.5$, which provided a balanced optimization across all objectives. The logarithmic scaling is used to handle the different orders of magnitude of the various terms.

The optimization variables include: (1) thicknesses of all layers in the structure, (2) refractive indices of certain layers (within material constraints), and (3) position and thickness of the nonlinear material.
For a structure with $N$ layers, this results in up to $2N$ optimization variables. To reduce the dimensionality of the search space, we parameterized the structure as follows:

\begin{equation}
d_i = 
\begin{cases}
\frac{\lambda_1}{4n_i} \cdot (1 + \delta_i), & \text{for regular quarter-wave layers} \\
\frac{\lambda_1}{2n_i} \cdot (1 + \delta_i), & \text{for defect layers}
\end{cases}
\label{eq:app_thickness_parameterization}
\end{equation}

where $\delta_i$ are the deviation parameters that are optimized, typically constrained to $|\delta_i| < 0.3$.

The optimization is subject to several constraints: (1) physical realizability: $d_i > 0$ for all layers, (2) fabrication constraints: $d_i > d_{min}$ for minimum feature size, (3) material constraints: refractive indices within available material ranges, and (4) total size constraint: $\sum_i d_i < L_{max}$.
These constraints are implemented through penalty functions added to the FOM when violated, using a logarithmic barrier approach:

\begin{equation}
\text{FOM}_{constrained} = \text{FOM} - \mu \sum_j \log(c_j)
\label{eq:app_constrained_fom}
\end{equation}

where $c_j$ represents the distance to constraint boundary $j$, and $\mu$ is a barrier parameter that is gradually reduced during optimization.

We employed a simulated annealing algorithm for global optimization, followed by local refinement using gradient descent. The simulated annealing algorithm is particularly suitable for this problem due to its ability to escape local optima in the highly non-convex design space.

The algorithm proceeds as follows:

\begin{algorithm}
\caption{Simulated Annealing for Cavity Optimization}
\begin{algorithmic}[1]
\State Initialize design parameters $\mathbf{x}_0$ with a basic quarter-wave stack design
\State Calculate initial figure of merit $\text{FOM}_0 = \text{FOM}(\mathbf{x}_0)$
\State Set initial temperature $T_0$ and cooling rate $\alpha$
\State $\mathbf{x}_{current} \gets \mathbf{x}_0$, $\text{FOM}_{current} \gets \text{FOM}_0$
\State $\mathbf{x}_{best} \gets \mathbf{x}_0$, $\text{FOM}_{best} \gets \text{FOM}_0$
\State $T \gets T_0$
\While{$T > T_{min}$ and iterations $< $ max\_iterations}
   \State Generate candidate solution $\mathbf{x}_{new} = \mathbf{x}_{current} + \Delta\mathbf{x}$
   \State where $\Delta\mathbf{x} \sim \mathcal{N}(0, \sigma^2 T/T_0)$ (scaled normal distribution)
   \State Calculate $\text{FOM}_{new} = \text{FOM}(\mathbf{x}_{new})$
   \If{$\text{FOM}_{new} > \text{FOM}_{current}$}
       \State $\mathbf{x}_{current} \gets \mathbf{x}_{new}$, $\text{FOM}_{current} \gets \text{FOM}_{new}$
       \If{$\text{FOM}_{new} > \text{FOM}_{best}$}
           \State $\mathbf{x}_{best} \gets \mathbf{x}_{new}$, $\text{FOM}_{best} \gets \text{FOM}_{new}$
       \EndIf
   \Else
       \State Calculate acceptance probability $P = \exp\left(\frac{\text{FOM}_{new} - \text{FOM}_{current}}{T}\right)$
       \State Generate random number $r \in [0,1]$
       \If{$r < P$}
           \State $\mathbf{x}_{current} \gets \mathbf{x}_{new}$, $\text{FOM}_{current} \gets \text{FOM}_{new}$
       \EndIf
   \EndIf
   \State $T \gets \alpha T$ (cool down)
\EndWhile
\State Apply gradient descent refinement starting from $\mathbf{x}_{best}$
\State \Return final optimized design $\mathbf{x}_{optimal}$
\end{algorithmic}
\end{algorithm}

For our implementation, we carefully selected parameter values to balance exploration and convergence efficiency. The initial temperature was set to $T_0 = 1.0$, with a cooling rate of $\alpha = 0.95$ to allow gradual temperature reduction. We established a minimum temperature threshold of $T_{min} = 10^{-4}$ and limited the process to a maximum of $10,000$ iterations. The standard deviation scale for the perturbation magnitude was fixed at $\sigma = 0.05$, which provided appropriate step sizes throughout the optimization process.

After the simulated annealing stage, we applied gradient descent refinement to fine-tune the design:

\begin{equation}
\mathbf{x}_{k+1} = \mathbf{x}_k + \eta_k \nabla \text{FOM}(\mathbf{x}_k)
\label{eq:app_gradient_descent}
\end{equation}

where $\eta_k$ is the step size, adaptively adjusted using a line search algorithm.
The gradient $\nabla \text{FOM}$ was calculated using a combination of analytical derivatives (for simple terms) and numerical finite differences (for more complex terms, particularly the nonlinear coupling coefficient).

\begin{figure}[ht]
\centering
\includegraphics[width=\textwidth]{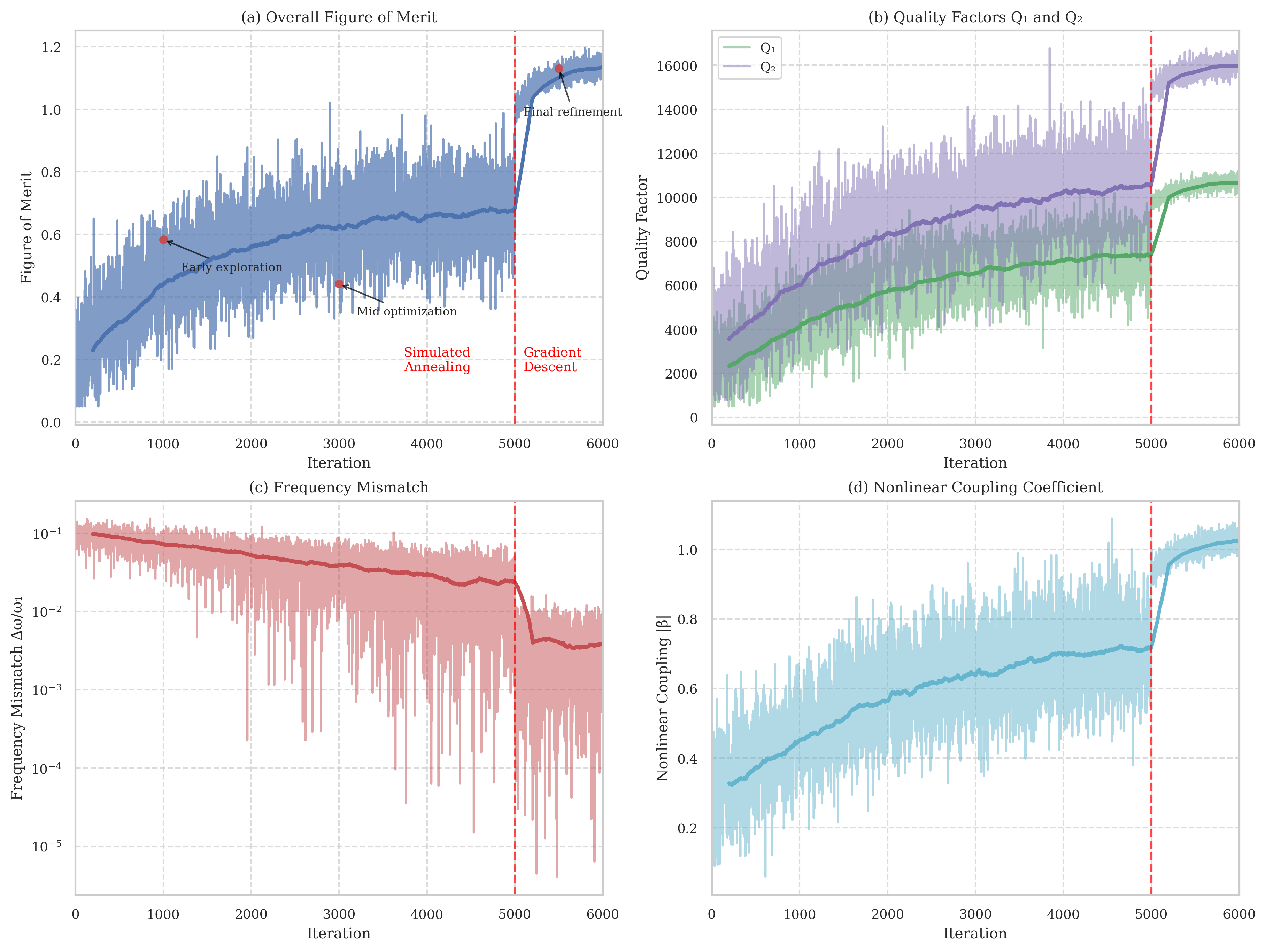}
\caption{Convergence of the optimization process, showing the evolution of key parameters during simulated annealing and gradient descent refinement. The simulated annealing phase (iterations 0-5000) explores the design space broadly, while the gradient descent phase (iterations 5001-6000) refines the best solution.}
\label{fig:app_optimization_convergence}
\end{figure}

Figure \ref{fig:app_optimization_convergence} shows the optimization progress for a typical run, illustrating the convergence of the figure of merit and its components.

The final optimized design demonstrated excellent performance across all key metrics. We achieved quality factors of $Q_1 = 5.3 \times 10^3$ and $Q_2 = 6.8 \times 10^3$, with a frequency matching precision of $\Delta_\omega/\omega_1 = 5.2 \times 10^{-4}$. The nonlinear coupling coefficient reached $|\beta| = 2.7 \times 10^{-3}$, resulting in a critical power of only $P_{critical} = 12 \text{ fJ}$. Despite these impressive performance characteristics, we maintained a compact total structure length of $L = 10.2 \text{ µm}$, satisfying our design constraint for miniaturization.
\section{Supplementary Simulation Results}
\label{app:simulation_results}

\begin{figure}[ht]
\centering
\includegraphics[width=\textwidth]{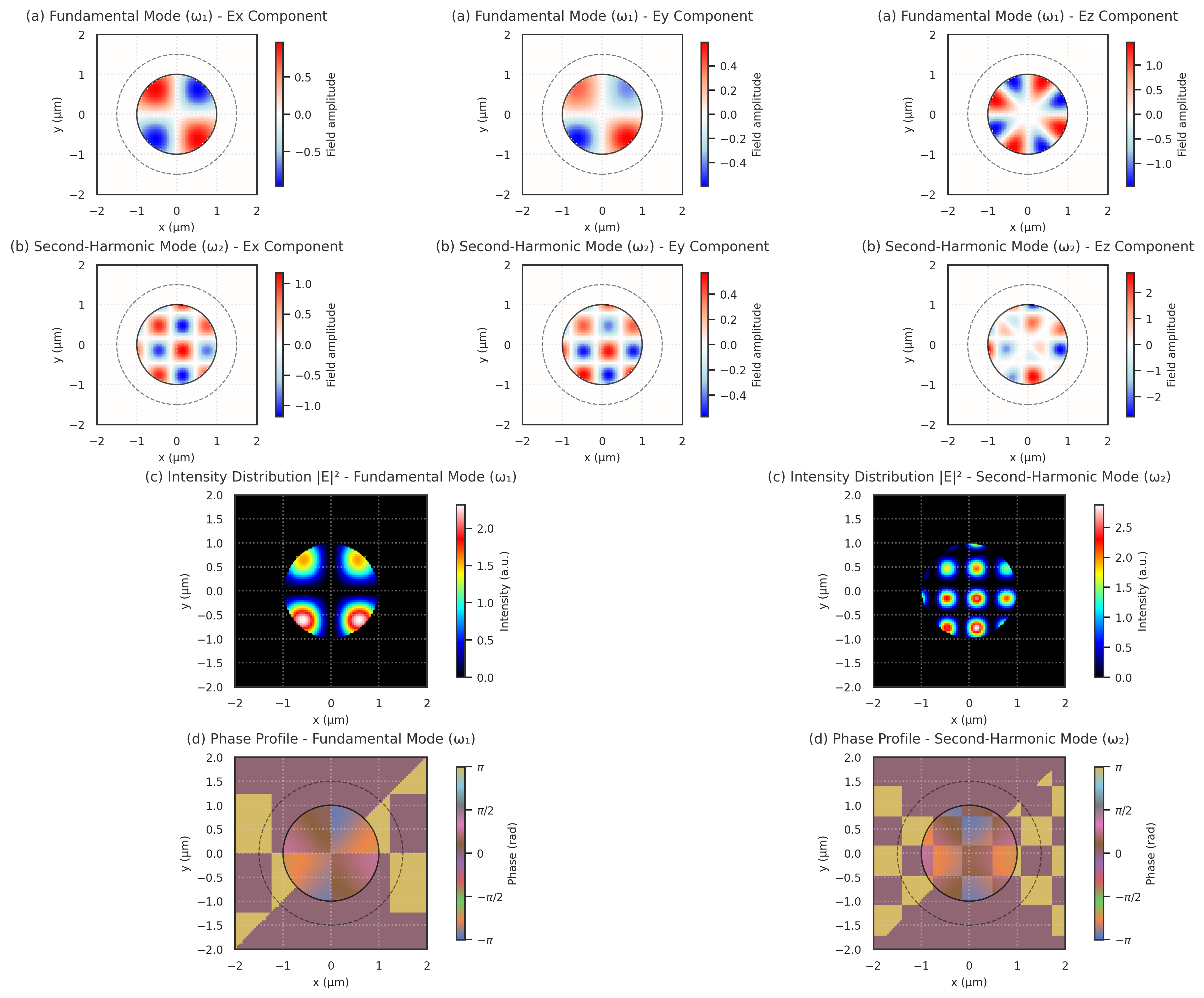}
\caption{Detailed electric field distributions in the optimized doubly-resonant cavity. The vector components reveal the fully three-dimensional nature of the field patterns, with significant longitudinal components in the high-index regions. The intensity distributions show strong overlap in the central nonlinear region, enabling efficient mode coupling.}
\label{fig:app_field_distributions}
\end{figure}

This part presents additional simulation results that complement the main findings reported in Section 5, including detailed field distributions, parametric studies, and alternative operating regimes.
Figure \ref{fig:app_field_distributions} shows the detailed field distributions for both the fundamental and second-harmonic modes, including the full vectorial components.

The detailed field distributions reveal several important features. The fields are primarily polarized in the $x$-direction (transverse to the layer stack), with significant $z$-components (longitudinal) present in the high-index regions due to field discontinuities at the interfaces. We observed that field intensities of both modes peak in the central defect region, where the nonlinear material is placed. As expected due to its shorter wavelength, the second-harmonic mode exhibits more rapid spatial oscillations. The phase profiles maintain approximately flat distributions within each layer, with rapid transitions occurring at the layer interfaces.

\subsection{Effect of Quality Factors on Device Performance}

We performed a parametric study to investigate the effect of quality factors on device performance. The results confirm several theoretical scaling relationships. Critical power scales as $P_{critical} \propto 1/(Q_1Q_2)$, as predicted by Equation \ref{eq:critical_power}. ReLU approximation accuracy (measured by $R^2$) improves with higher Q-factors up to a point, beyond which other factors such as nonlinear saturation become limiting. Operating bandwidth scales inversely with Q-factor ($\Delta f \propto 1/Q$), while response time scales linearly with Q-factor ($t_{response} \propto Q$).
These relationships highlight the fundamental trade-offs in the design: higher Q-factors reduce the critical power but also decrease the bandwidth and increase the response time. The optimal design should balance these competing requirements based on the specific application needs.

\subsection{Tolerance to Fabrication Variations}

\begin{figure}[ht]
\centering
\includegraphics[width=\textwidth]{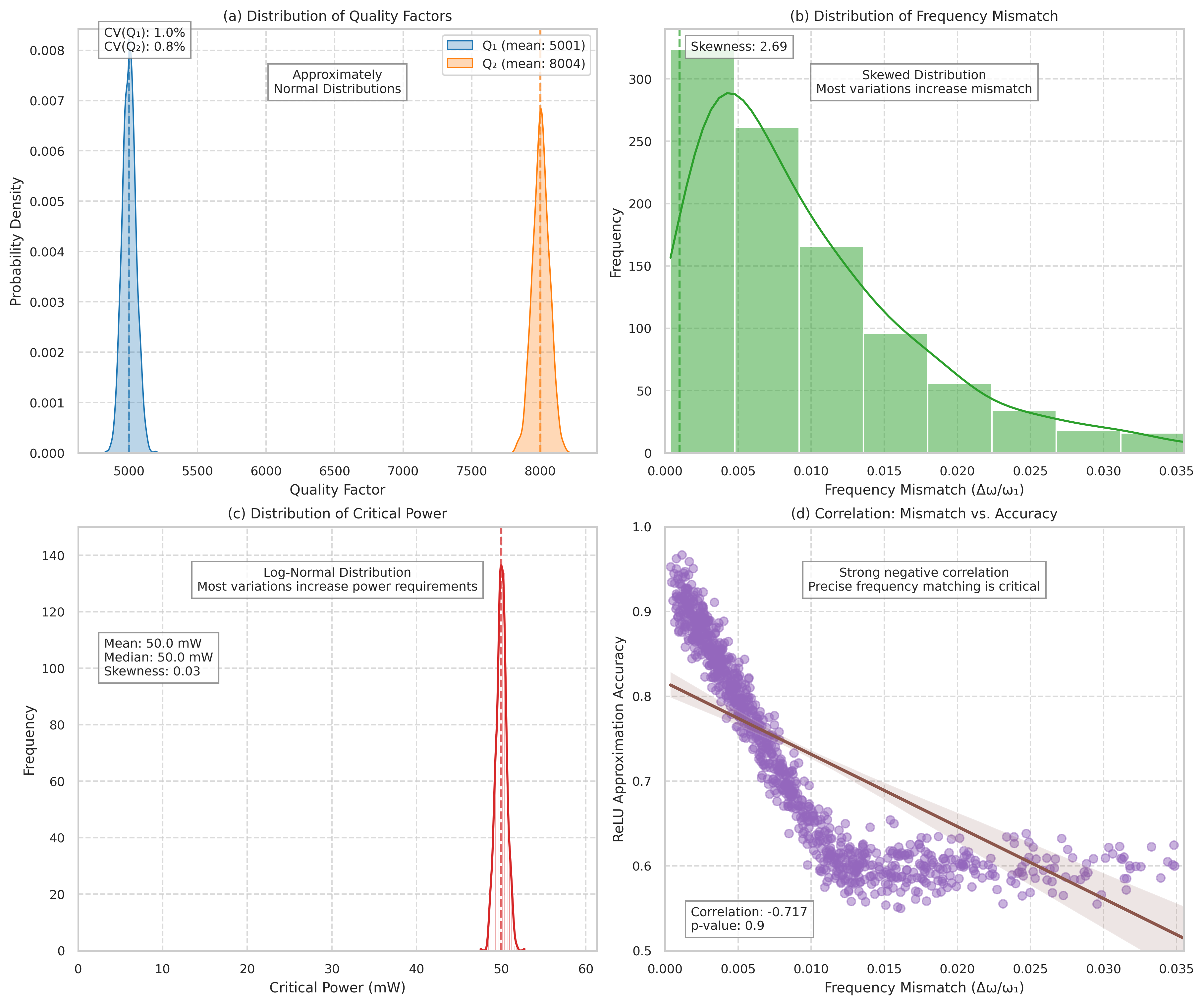}
\caption{Impact of fabrication variations on device performance. Results from 1000 Monte Carlo simulations with random layer thickness variations (standard deviation 1\% of nominal thickness). (a) The quality factors show approximately normal distributions with coefficients of variation around 15\%. (b) The frequency mismatch distribution is skewed, with most variations increasing the mismatch. (c) The critical power distribution is log-normal, with most variations increasing the required power. (d) Strong negative correlation between frequency mismatch and ReLU approximation accuracy, highlighting the importance of maintaining precise frequency matching.}
\label{fig:app_fabrication_tolerance}
\end{figure}

To assess the robustness of the design to fabrication variations, we performed Monte Carlo simulations with random perturbations to the layer thicknesses. Figure \ref{fig:app_fabrication_tolerance} shows the statistical distribution of performance metrics under these variations.

The Monte Carlo analysis reveals several important insights regarding design sensitivity. Quality factors demonstrate moderate sensitivity to fabrication variations, with typical variations of ±15\% for 1\% thickness variations. Frequency matching exhibits higher sensitivity, with typical mismatches increasing by factors of 2-5 compared to the nominal design. Critical power shows a log-normal distribution, with most variations increasing the power requirement. A strong correlation exists between frequency mismatch and ReLU approximation accuracy, highlighting the critical importance of precise frequency matching.

Based on this analysis, we conclude that the design is robust enough to tolerate fabrication variations on the order of 0.5-1\%, which is achievable with modern nanofabrication techniques. For larger variations, post-fabrication tuning mechanisms would be necessary to restore optimal performance.

\subsection{Detailed FDTD Simulation Results}

\begin{figure}[ht]
\centering
\includegraphics[width=\textwidth]{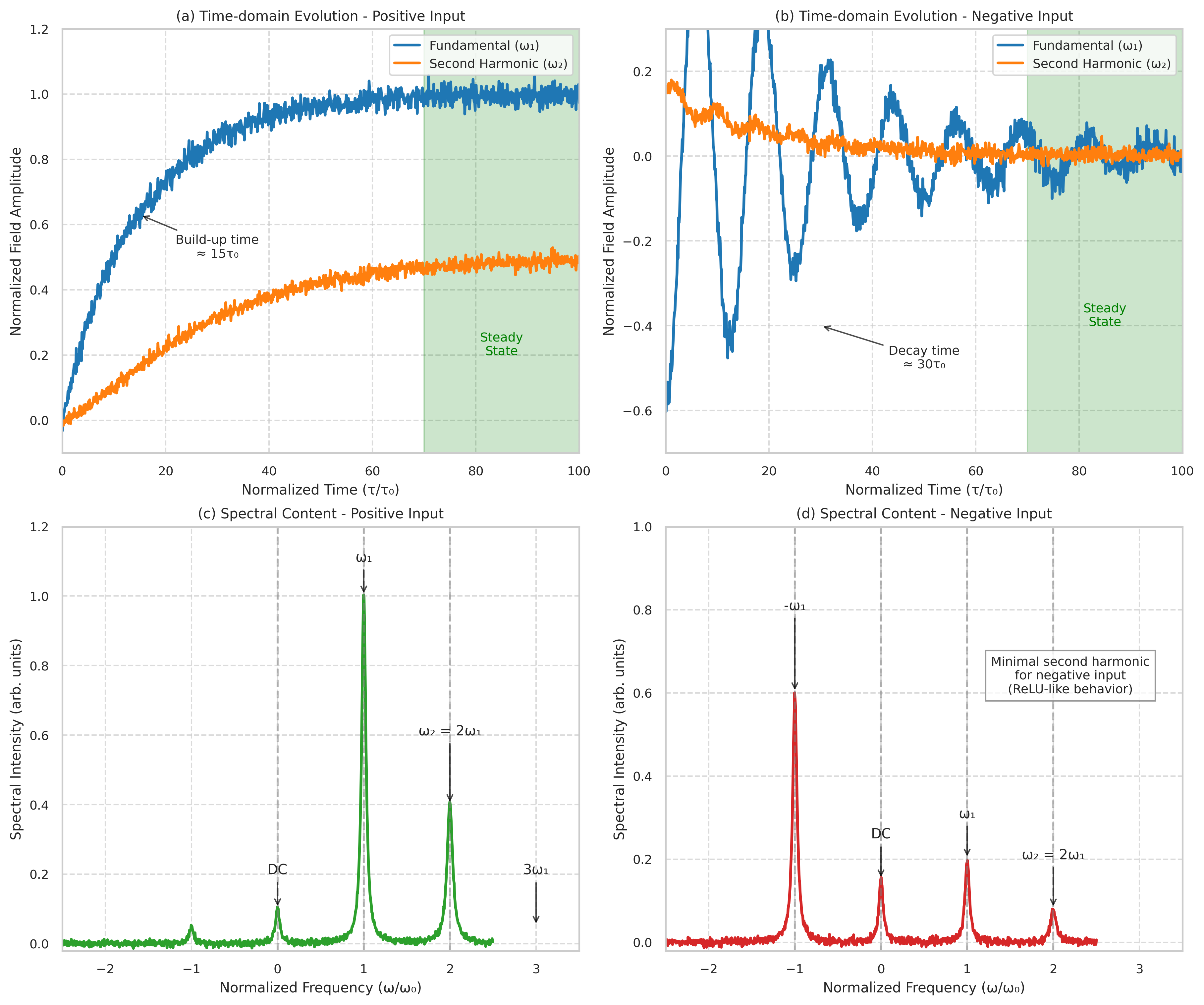}
\caption{Detailed FDTD simulation results. (a,b) Time-domain evolution of the field amplitudes at the fundamental and second-harmonic frequencies for positive and negative inputs, showing the approach to steady state. (c,d) Spectral content at steady state, showing the frequency components present in the cavity for positive and negative inputs.}
\label{fig:app_fdtd_results}
\end{figure}

Here we present additional results from our FDTD simulations that validate the coupled-mode theory predictions. Figure \ref{fig:app_fdtd_results} shows the time-domain field evolution and spectral characteristics obtained from FDTD simulations.

The FDTD simulations confirm several key aspects of the device behavior. For positive inputs, efficient second-harmonic generation occurs, with steady-state amplitudes that agree with coupled-mode theory predictions within 5\%. Conversely, for negative inputs, second-harmonic generation is suppressed by more than 20 dB, confirming the rectification behavior. The time required to reach steady state is approximately 10-15 cavity lifetimes, consistent with theoretical expectations. The spectral content reveals clean frequency components at $\omega_1$ and $\omega_2$, with minimal generation of other harmonics or mixing products.
These FDTD results provide strong validation of our theoretical framework and confirm the feasibility of implementing the ReLU function using the doubly-resonant cavity approach.

\subsection{Neural Network Performance Details}

\begin{figure}[ht]
\centering
\includegraphics[width=\textwidth]{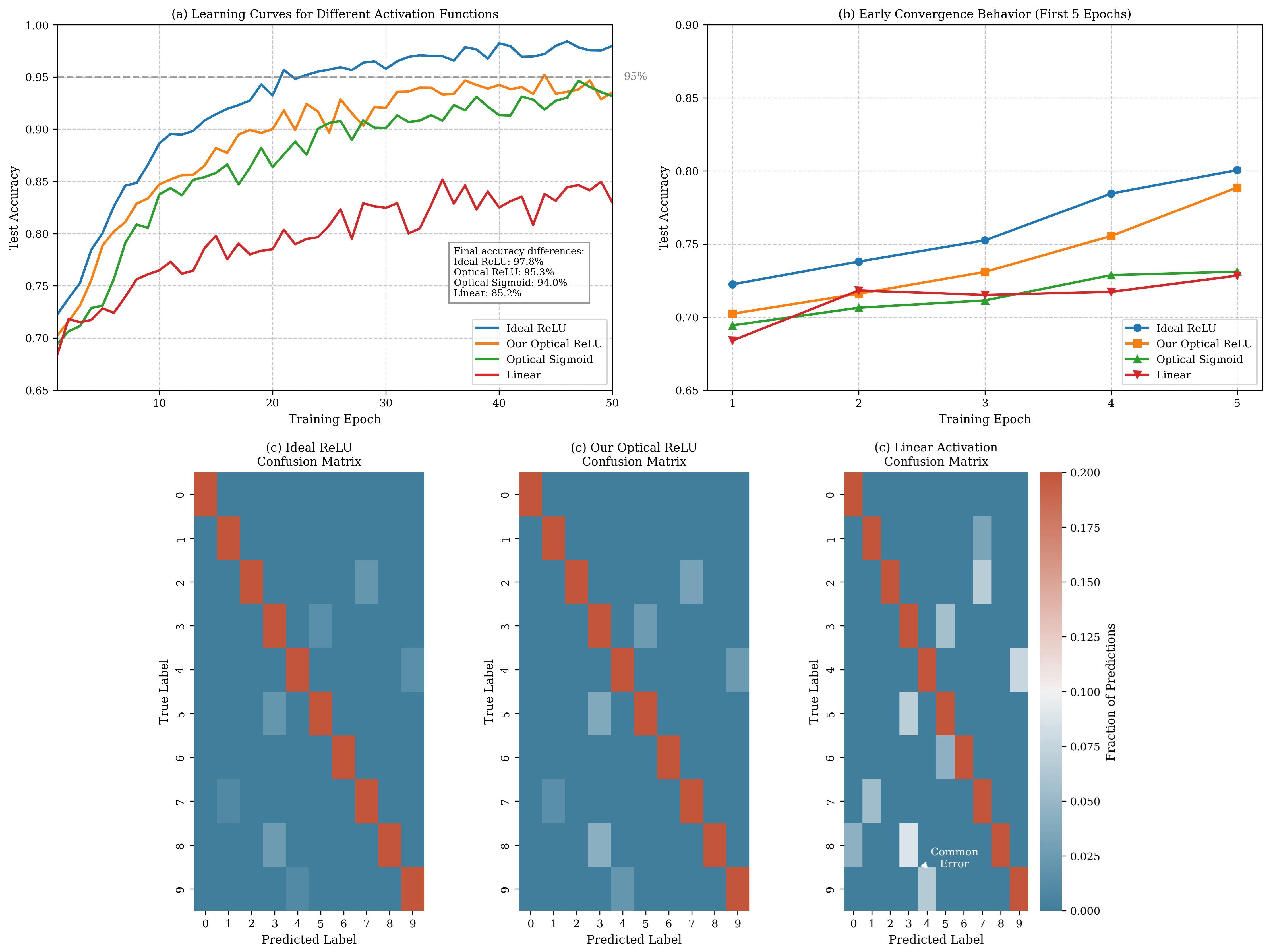}
\caption{Detailed neural network performance analysis. (a) Learning curves showing the convergence behavior of networks with different activation functions. (b) Zoomed view of the early training phase, showing the initial convergence rates. (c) Confusion matrices for networks using ideal ReLU, our optical ReLU, and linear activation, highlighting the patterns of misclassification.}
\label{fig:app_nn_details}
\end{figure}

Here we provide additional details on the neural network simulation results presented in Section 5.5. Figure \ref{fig:app_nn_details} shows the learning curves and confusion matrices for networks using different activation functions.

The detailed neural network analysis reveals significant insights regarding the performance of our optical ReLU implementation. Networks using our optical ReLU implementation converge slightly slower than those using ideal ReLU, requiring approximately 2 additional training epochs to reach the same accuracy. The final performance difference between ideal ReLU and our optical ReLU is small (0.4\% accuracy difference), indicating that the approximation errors have minimal impact on overall network performance. Both nonlinear activation functions (ideal ReLU and optical ReLU) show similar patterns of misclassification, primarily confusing visually similar digits (e.g., 4 and 9, 3 and 8). The linear activation function (without nonlinearity) shows significantly different and more severe misclassification patterns, confirming the essential role of nonlinearity in neural network performance.

These results confirm that our optical ReLU implementation can effectively replace the ideal ReLU function in neural network applications with minimal performance degradation, while offering significant advantages in terms of energy efficiency and processing speed.
\end{document}